\pdfoutput=1 

\documentclass[sigplan,10pt,nonacm,balance=false]{acmart}
\setcopyright{none}

\usepackage[utf8]{inputenc}
\usepackage[T1]{fontenc}
\usepackage{subcaption}
\usepackage{graphicx}    
\usepackage{multirow}
\usepackage{array}
\usepackage{tabularx} 
\usepackage{booktabs} 
\usepackage{makecell}
\usepackage{xspace}
\usepackage[linesnumbered,ruled,vlined]{algorithm2e}
\usepackage{amsmath}
\usepackage{amssymb}
\usepackage{enumitem}
\usepackage{mathtools}
\usepackage{amsthm}
\usepackage{colortbl}
\usepackage[dvipsnames]{xcolor}
\usepackage{wrapfig}
\usepackage{textcomp}
\usepackage{siunitx}
\usepackage{caption}
\usepackage{tikz}
\usepackage{bm}
\usepackage{threeparttable}

\theoremstyle{plain}

\theoremstyle{definition}

\theoremstyle{remark}

\newcommand{\OURS}{\textsc{HyDra}\xspace}

\ifdefined\pdfmapline\pdfmapline{+txhy <T1-WGL4.enc <txhy.ttf}\fi
\DeclareFontFamily{T1}{txhy}{}
\DeclareFontShape{T1}{txhy}{m}{n}{<-> s*[0.9] txhy}{}
\newcommand{\txhy}{\fontencoding{T1}\fontfamily{txhy}\fontseries{m}\fontshape{n}\selectfont}

\newcommand{\Insight}[2]{%
  \vspace{3mm}%
  \setlength{\fboxsep}{6pt}
  \noindent\fcolorbox{black}{gray!10}{%
    \begin{minipage}{0.94\linewidth}
      \itshape \textbf{Insight #1:} #2
    \end{minipage}%
  }%
  \vspace{2mm}%
}

\begin{document}



\title{{\txhy\textcolor{blue}{Hy}}\textit{Dra}: Demystifying and Taming Dynamic Context Parallelism at Production Scale}

\newcommand{\sjtu}{\dag}    
\newcommand{\hy}{\ddag}
\newcommand{\nju}{\S}
\newcommand{\corresponding}{\P}

\author{Zihao Fan$^{\sjtu\hy}$, Yunzhuo Liu$^{\hy}$, Bo Jiang$^{\sjtu}$, Changgang Zheng$^{\nju}$, Lin Zheng$^{\hy}$, Ray Ying$^{\hy}$, Key Zhang$^{\hy}$
\\$^{\sjtu}$Shanghai Jiao Tong University \quad $^{\hy}$Tencent Hy \quad $^{\nju}$Nanjing University}
\renewcommand{\shortauthors}{Fan et al.}


\renewcommand\footnotetextcopyrightpermission[1]{} 

\newcommand*\varcircled[1]{\raisebox{.5pt}{\textcircled{\raisebox{-0.2pt}{\hspace{-.5pt}\scriptsize #1}}}}

\begin{abstract}
Long-context training runs on sequences whose lengths span orders of magnitude, and dynamic context parallelism (DCP) gives each sequence its own CP degree.
Existing DCP systems either do not scale or perform poorly on mainstream models, leaving Megatron-Core (Mcore) DCP as the only option at production scale.
Mcore DCP, however, sizes each degree to fit memory, which grows linearly with length while attention grows quadratically, so comparable token counts hide unequal computation.
In our production 256K-context training job on more than \textbf{11K} GPUs under Mcore DCP, per-rank microbatch times differ by up to $5\times$ at similar token counts.
The skew leads to a $46\%$ pipeline bubble and a $13\%$ data-parallel bubble.
We present \OURS, a scalable load-driven DCP system.
Its scheduler balances computation by pulling every rank toward one load target, and balance in turn makes that target solvable in closed form.
It places sequences with lazy heaps rather than whole-pool scans.
That balance asks for CP degrees larger than memory requires, so its CP engine nests an inner Ulysses group in a shallow outer ring, letting a higher degree lower computation and communication together.
Evaluation at both scales shows consistent gains.
On a \textbf{512}-GPU testbed, \OURS raises throughput over Mcore DCP by $1.18\times$ on average at 32K context and $2.48\times$ at 256K.
On a \textbf{2,048}-GPU production job, it shrinks the pipeline bubble from $36\%$ to $14\%$, cuts scheduling time by $2.3\times$, and raises throughput by $1.10$--$1.43\times$ (avg. $1.25\times$) over Mcore DCP and $1.33$--$1.90\times$ (avg. $1.59\times$) over static CP.
\end{abstract}

\newcommand{\bj}[1]{\textcolor{red}{(BJ:#1)}}
\maketitle

\section{Introduction}
\label{sec:introduction}

Long-context capabilities are increasingly important for frontier language models~\cite{dai2019transformerxl, beltagy2020longformer, zaheer2020bigbird, liu2023blockwise, dao2022flashattention, dao2024flashattention2, grattafiori2024llama3}.
The data used for long-context training, however, do not form a uniform stream of long sequences. 
Real corpora mix short documents with a small fraction of very long ones, so sequence lengths span orders of magnitude~\cite{ge2025bytescale, wang2025flexsp, wang2025wlbllm}.
Dynamic context parallelism (DCP) is an effective way to train on such data, because it adapts the context-parallel (CP) degree~\cite{liu2023ring, jacobs2023ulysses} to each sequence instead of fixing it for the whole job~\cite{wang2025flexsp, ge2025bytescale}.

In practice, however, every existing DCP system falls short in a different way~\cite{ge2025bytescale, wang2025flexsp, li2025hydraulis, jiang2025dcp, nvidia2025dynamiccp}.
The first limitation is scalability.
FlexSP~\cite{wang2025flexsp}, Hydraulis~\cite{li2025hydraulis}, and DCP~\cite{jiang2025dcp} all plan every iteration by solving a global optimization problem, which becomes prohibitively expensive at large scale.
The second limitation is performance.
ByteScale~\cite{ge2025bytescale} does scale, but its ring-based CP~\cite{liu2023ring} is designed for grouped-query attention (GQA)~\cite{ainslie2023gqa}, where the few KV heads keep the exchange cheap.
Mainstream models now use multi-head latent attention (MLA)~\cite{deepseekai2024deepseekv2,liu2024deepseekv3,deepseek2026v4}, which materializes K and V for every head during training, so the same exchange moves far more data and performance degrades sharply at large scale.

What remains is the DCP shipped in Megatron-Core (Mcore)~\cite{nvidia2025dynamiccp}, which is today the only implementation usable at production scale.
Its scheduler, however, sizes the CP group of each sample from a memory budget, namely the largest number of tokens a rank can hold.
Memory grows linearly with sequence length while attention computation grows quadratically, so a CP size chosen to fit memory leaves the computation imbalanced, and the imbalance worsens as sequences get longer.
We confirm this on a 256K-context training job, running Mcore DCP across more than \textbf{11K} GPUs in our model production (\S\ref{sec:measurement}).
Per-rank token counts are comparable, yet microbatch execution times differ by $5\times$ and CP-group loads by $2.4\times$.
This imbalance leads to a $46.0\%$ pipeline (PP) bubble and a data-parallel (DP) bubble that adds up to $12.8\%$ to every iteration.
Planning costs another $9\%$, because the scheduler rescans the entire pool of sequences on every iteration.

We present \OURS, a load-driven DCP system that keeps both planning and execution efficient at production scale.
\textbf{(1) For execution}, \OURS derives one per-rank load target for the whole iteration from the quadratic attention computation, and sizes every sequence's CP degree to meet it, pulling every rank toward the same load and shrinking both the PP and DP bubble.
The target is the lowest one the longest sequence can meet, which also makes it the easiest for short sequences to reach.
Long sequences reach that target only at large CP degrees, which far exceed the upper bound of attention-head count in Ulysses.
Ring does support such large CP degrees, but its per-rank traffic does not shrink as the degree grows, so communication becomes the bottleneck, preventing long sequences from reaching the target.
\OURS therefore factorizes each CP degree into an inner Ulysses group inside a shallow outer ring, and overlaps the exposed all-to-alls with computation.
A higher degree now lowers both computation and communication per rank, so even the longest sequences can reach the load target.
\textbf{(2) For planning}, \OURS derives these quantities in closed form and places sequences with heaps.
Iteration time is a graph of waits, each a maximum over ranks, so it has no closed form.
Balance removes those maxima and leaves one unknown, the largest CP degree allowed, in a convex trade-off between the pipeline bubble and the per-MB overhead.
\OURS therefore solves that trade-off in closed form, and the load target and MB count follow from the degree by substitution.
Placement is the only cost left, and a lazy heap per CP degree turns its whole-pool scans into logarithmic lookups.


We implement \OURS in approximately 3K LoC of Python on Megatron-LM~\cite{shoeybi2019megatron}.
On a \textbf{512}-GPU testbed, it improves end-to-end throughput over Mcore DCP by $1.09$--$1.25\times$ (avg. $1.18\times$) at 32K context and $1.42$--$3.08\times$ (avg. $2.48\times$) at 256K, keeping the compute stream busy over $76\%$ of each iteration at both lengths.
In production 256K-context training on \textbf{2,048} GPUs, it improves throughput by $1.10$--$1.43\times$ (avg. $1.25\times$) over Mcore DCP and $1.33$--$1.90\times$ (avg. $1.59\times$) over static CP.
It shrinks the pipeline bubble from $36.4\%$ to $13.6\%$ of each iteration and cuts scheduling time by $2.3\times$, rising to $11.6\times$ at 16K GPUs, while its training-loss curves track static CP over roughly $8{,}000$ iterations.
In simulation, it averages $1.28\times$ over the state-of-the-art (SOTA) ByteScale~\cite{ge2025bytescale} on a GQA model and $1.53\times$ on a communication-heavy MLA model.

This paper makes the following contributions:
\begin{itemize}[leftmargin=*,nosep]
    \item We characterize production Mcore DCP on more than \textbf{11K} GPUs, where up to a $46.0\%$ PP bubble, a $12.8\%$ DP bubble, and a $9\%$ scheduling stall dominate: balancing tokens instead of attention leaves MB times $5\times$ and CP-group loads $2.4\times$ apart.
    
    \item We propose \OURS, a load-driven DCP system that collapses the PP and DP bubbles with one load target. 
    \OURS co-designs planning with execution, so the CP degrees that target demands stay affordable at production scale.
    
    \item We design a scheduler that derives the load target in closed form and places sequences with lazy heaps rather than whole-pool scans. 
    We break the ring--Ulysses dilemma by running a Ulysses group inside a shallow ring, so a larger degree lowers computation and communication together.
    
    \item We implement \OURS in Megatron-LM and evaluate it at 32K and 256K context: on a \textbf{512}-GPU testbed it raises throughput over Mcore DCP by $1.09$--$1.25\times$ (avg. $1.18\times$) at 32K and $1.42$--$3.08\times$ (avg. $2.48\times$) at 256K, on \textbf{2,048} production GPUs by $1.10$--$1.43\times$ (avg. $1.25\times$) over Mcore DCP and $1.33$--$1.90\times$ (avg. $1.59\times$) over static CP while cutting scheduling time by $2.3\times$; in simulation to 40K GPUs it leads the SOTA baseline by $1.28$--$1.53\times$.
\end{itemize}

\section{Background}
\label{sec:background}

\subsection{5D Parallelism for Distributed LLM Training}

Large-scale LLM training spans five parallelism dimensions, which planners combine under model, memory, and topology constraints~\cite{shoeybi2019megatron, narayanan2021efficient, jiang2024megascale, jia2019flexflow, zheng2022alpa}.
Tensor parallelism (TP) shards operators within a node~\cite{shazeer2018meshtensorflow, shoeybi2019megatron}; pipeline parallelism (PP) splits layers into microbatched stages, trading bubbles against activation memory~\cite{huang2019gpipe, narayanan2019pipedream}; expert parallelism (EP) shards mixture-of-experts (MoE) weights and routes tokens through all-to-all~\cite{lepikhin2020gshard, fedus2022switch}; and data parallelism (DP) replicates the model, shards the batch, and synchronizes gradients~\cite{li2020pytorch, rajbhandari2020zero}.
Context parallelism (CP) shards each sequence across ranks, so each rank holds only a fraction of its activations and long contexts become feasible~\cite{li2023sequenceparallelism, liu2023ring, jacobs2023ulysses, li2024distflashattn, fang2024usp}.
Because attention couples all tokens, CP circulates key and value blocks around a ring or switches sequence and head shardings through all-to-all~\cite{liu2023ring,jacobs2023ulysses}.
This work fixes TP, PP, and EP at their deployment settings and jointly manages CP and DP to place variable-length sequences across ranks.

\begin{figure}[t]
    \centering
    \includegraphics[width=0.85\linewidth]{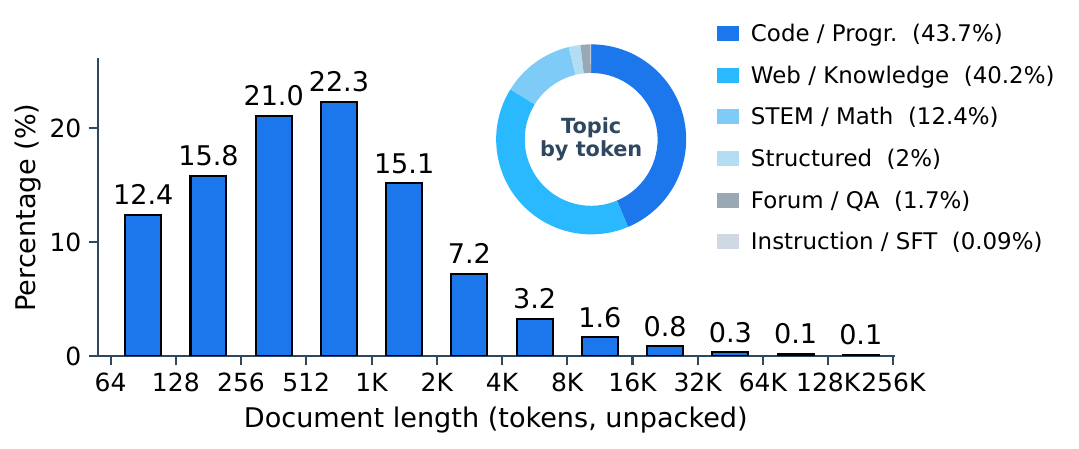}
    \caption{\textit{Profile of the 256K-context training dataset.}}
    \label{fig:dataset_profile_256k}
\end{figure}

\subsection{Long-Context Data Characteristics}
Long-context corpora have highly skewed sequence-length distributions~\cite{wang2025flexsp, ge2025bytescale}.
In the benchmark dataset used for our measurements (Figure~\ref{fig:dataset_profile_256k}), lengths span four orders of magnitude up to 256K tokens, with more than 85\% of sequences below 2K and fewer than 0.6\% above 32K.
The rare long ones are what teach long-context ability, so every batch mixes them with abundant short ones~\cite{grattafiori2024llama3}.

To fill each context window, systems pack short sequences together and mask attention across their boundaries~\cite{krell2021packing, ding2024packing, staniszewski2025splice}, while ragged-tensor compilation and mini-sequence execution cut padding and non-attention memory further~\cite{fegade2022cora, luo2024minisequence}.
Packing equalizes token counts, not attention work, which scales as $\sum_i \mathcal{O}(s_i^2)$, so packed samples of the same size can differ widely in computation.

\begin{figure}[b]
    \centering
    \includegraphics[width=0.95\linewidth]{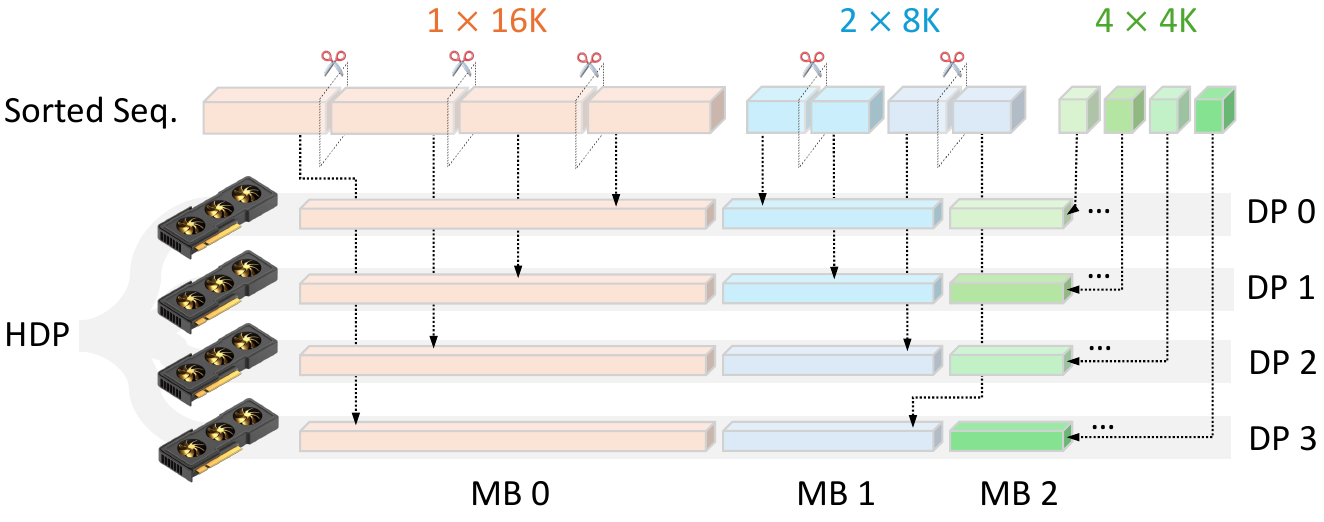}
    \caption{\textit{DCP under a 4K per-rank token budget.}}
    \label{fig:dcp}
\end{figure}

\begin{figure*}[t]
    \centering
    \includegraphics[width=1\textwidth]{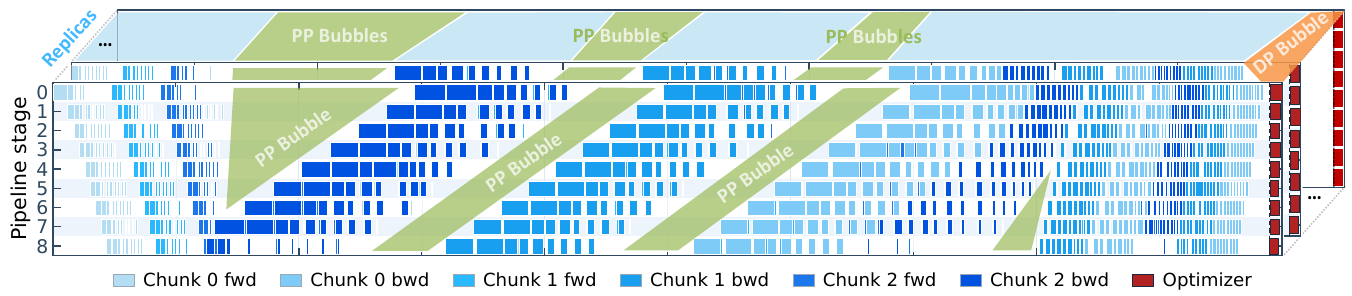}
    \caption{\textit{Training timeline of one iteration under Mcore DCP, across nine PP stages and DP replicas.}}
    \label{fig:pipeline_performance}
\end{figure*}

\subsection{Dynamic Context Parallelism}

DCP gives every sequence its own CP degree instead of one degree for the whole job~\cite{ge2025bytescale, wang2025flexsp, jiang2025dcp}.
Mcore ships a production scheduler that sizes each degree from a token budget, the most tokens a rank can hold, so every sequence goes to the fewest ranks that fit it~\cite{nvidia2025dynamiccp}.
Figure~\ref{fig:dcp} shows the rule on four ranks with a 4K budget, where a 16K sequence spans all four, an 8K sequence takes two, and a 4K sequence stays on one.
Short sequences thus avoid the redundant communication a job-wide degree would force on them.

CP and DP partition the same batch in two directions, CP within one sequence and DP across sequences, so a static job organizes its ranks as a DP$\times$CP mesh sized for the longest sequence.
Per-sequence degrees leave no fixed mesh, so the DP$\times$CP ranks become one pool, the \emph{hybrid data-parallel} (HDP) domain, from which each group is cut at placement time~\cite{ge2025bytescale, nvidia2025dynamiccp}.
In Figure~\ref{fig:dcp}, the same four ranks serve as one group of four in MB~0, as two groups of two in MB~1, and as four groups of one in MB~2.
In our production training, Mcore DCP improves throughput over static CP by about $8\%$ on 32K data and $39\%$ on 256K data.
The budget, however, counts tokens, and Section~\ref{sec:measurement} shows that this leaves quadratic attention work badly imbalanced.

\section{Characterizing DCP at Production Scale}
\label{sec:measurement}

In this section, we measure Mcore DCP at production scale and identify its major bottlenecks.

\subsection{Measurement Setup and Methodology}

\looseness=-1 All measurements in this section are collected from a training job of Hy4-preview~\cite{hy4preview}, an open-weight MoE model with 770B total and 49B activated parameters, running on more than \textbf{11K} GPUs at 256K context length.
The model is parallelized with $tp=2, pp=9, ep=32$, while the remaining ranks form the DP and CP dimensions managed by Mcore DCP~\cite{nvidia2025dynamiccp}, whose CP communication uses Ulysses-style all-to-all~\cite{jacobs2023ulysses}.
Traces are collected with \textsc{Argus}~\cite{zhou2026argus}, a lightweight, always-on tracing system.

\subsection{Inter-MB Imbalance}
\label{subsec:intermb}

Figure~\ref{fig:pipeline_performance} plots one iteration across all nine pipeline stages, with the forward and backward passes of every MB over three interleaved chunks and the closing optimizer step.

\textbf{The heaviest MB bounds the iteration.}
Iteration time follows the \emph{critical path} through this schedule: every stage must process every MB, so the longest-running one recurs stage after stage and dictates the iteration latency.
Figure~\ref{fig:pipeline_performance} traces one iteration of this job, where the heaviest MB, the first in the schedule, stretches every stage and idle time fans out around it.
Under DCP each MB packs a different mix of variable-length sequences at a different CP degree, and in this example the heaviest takes more than $5\times$ as long to execute as the lightest.
We call this skew \emph{inter-MB imbalance}; its footprint is the \emph{PP bubble}, the green gaps in the figure, which consume $45.97\%$ of this pipeline, nearly half of every iteration.
A balanced schedule at $pp{=}9$ with three interleaved chunks would bubble only about $10\%$ at this MB count, so skew inflates it $4.6\times$.

\begin{figure}[t]
    \centering
    \includegraphics[width=1\linewidth]{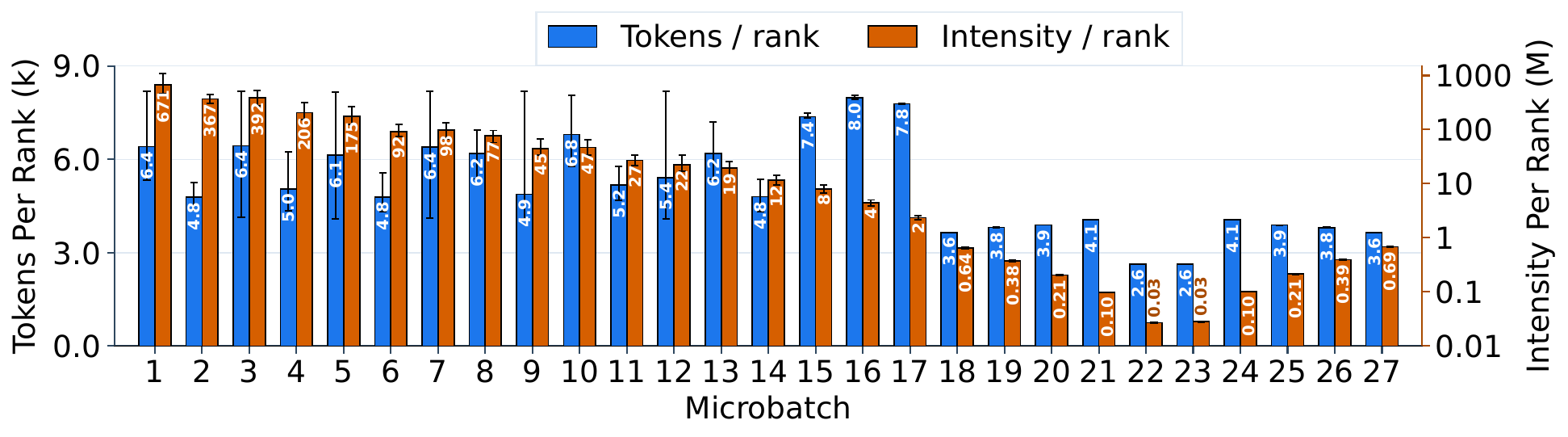}
    \caption{\textit{Per-rank profile of each MB.}}
    \label{fig:mb_workload_profile}
\end{figure}

\begin{figure*}[t]
    \centering
    \hspace{0.1em}\begin{minipage}[b]{0.3\linewidth}
        \centering
        \includegraphics[width=0.95\linewidth]{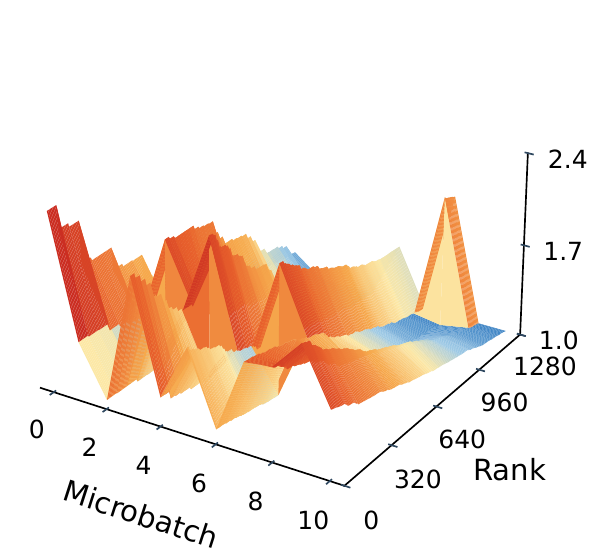}
        \caption{\textit{Inter-CP group imbalance.}}
        \label{fig:inter_cp_group_imbalance}
    \end{minipage}
    \hfill
    \nextfloat
    \begin{minipage}[b]{0.68\linewidth}
        \centering
        \hspace{-0.5em}\begin{subfigure}[b]{0.5\linewidth}
            \centering
            \includegraphics[width=1.03\linewidth]{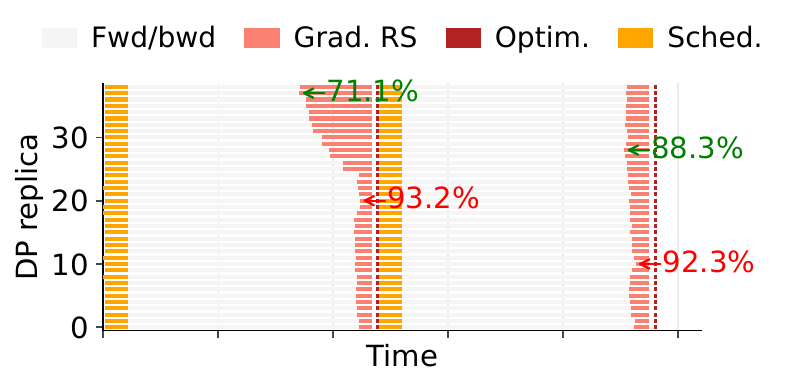}
            \subcaption{\textit{Per-iteration DP timeline}}
            \label{fig:dp_imbalance_iters}
        \end{subfigure}
        \hfill
        \begin{subfigure}[b]{0.24\linewidth}
            \centering
            \includegraphics[width=\linewidth]{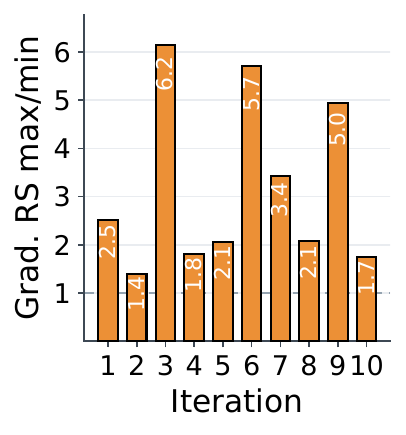}
            \subcaption{\textit{Grad. RS max/min}}
            \label{fig:dp_grad_maxmin}
        \end{subfigure}
        \hfill
        \begin{subfigure}[b]{0.24\linewidth}
            \centering
            \includegraphics[width=\linewidth]{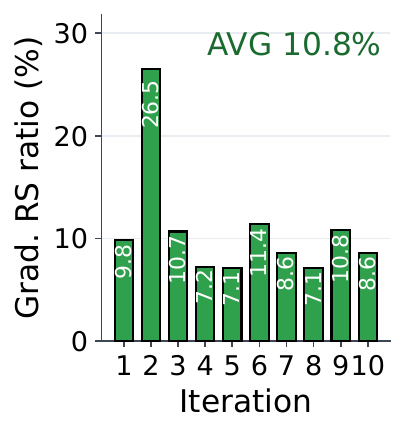}
            \subcaption{\textit{Grad. RS ratio}}
            \label{fig:dp_grad_ratio}
        \end{subfigure}
        \caption{\textit{DP bubble.}}
        \label{fig:dp_imbalance}
    \end{minipage}
\end{figure*}

The imbalance traces to a mismatch between how load is partitioned and how it scales.
Figure~\ref{fig:mb_workload_profile} profiles each MB in this iteration with two per-rank quantities.
The first is the token count, which grows linearly with length and is what the scheduler equalizes.
The second is the \emph{attention intensity} of a rank, $\mathcal{I}=\sum_i s_i^2/R$, the quadratic work of the sequences $\{s_i\}$ in the MB divided over the $R$ ranks that carry them.
The two diverge sharply.
Token counts stay of the same order, averaging around 5K, while intensity spans four orders of magnitude, from ${\approx}0.1$\,M to ${\approx}1{,}000$\,M.
Equal tokens need not imply equal computation, which has motivated FLOP-aware packing and scheduling~\cite{ge2025bytescale, wang2025wlbllm}; our point is that the skew survives \emph{production} DCP, whose scheduler equalizes tokens and memory per rank but never the quadratic attention cost.

\Insight{1}{\looseness=-1 Under Mcore DCP, per-rank tokens are balanced yet attention computation intensity varies by orders of magnitude, opening a $5\times$ MB execution-time gap and widening the PP bubble from its $10\%$ floor to $45.97\%$.
MBs must therefore be balanced by attention computation, not tokens alone.}

\subsection{Intra-MB Imbalance}
\label{subsec:intramb}

\looseness=-1 Imbalance also arises \emph{within} a single MB.
It starts as a computation load gap between the CP groups that share the MB, and every MoE all-to-all stamps that gap onto whole EP groups.
Nothing reconciles those groups until the iteration ends, so the gap may accumulate MB after MB into the \emph{DP bubble}.

\textbf{Inter-CP-group imbalance.}
An HDP domain consists of several CP groups, each responsible for a subset of an MB.
For each CP group we calculate the computation of the sequences it holds, then normalize these per-group values within each MB to its least-loaded group (set to $1$); Figure~\ref{fig:inter_cp_group_imbalance} plots the result for a representative subset of an iteration's MBs.
The spread is large: within the same MB, the busiest CP group carries up to $2.4\times$ the computation of the idlest.
It is worst for MB~0, which is also the MB that paces the iteration, so both forms of imbalance land on the same MB.

\begin{figure}[t]
    \centering
    \includegraphics[width=0.98\linewidth]{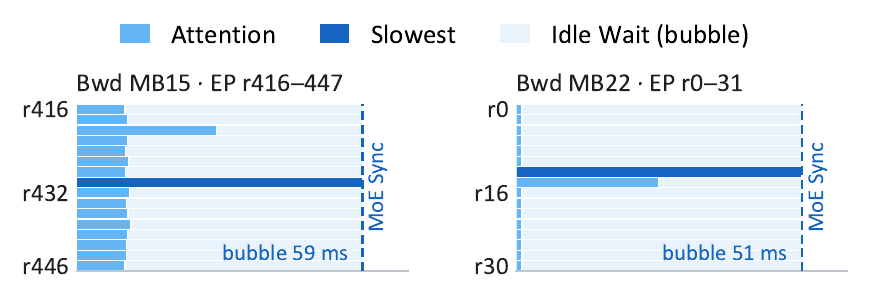}
    \caption{\textit{Intra-EP group imbalance.}}
    \label{fig:intra_ep_group_imbalance}
\end{figure}

\textbf{Intra-EP-group imbalance.}
Every MoE layer opens with a dispatch all-to-all, where the CP groups sharing an EP group must rendezvous before any of them can proceed.
Such a barrier cannot absorb the skew; it spreads it, stamping the slowest group's cost onto every group in the EP domain.
In Figure~\ref{fig:intra_ep_group_imbalance}, one rank's attention runs $70$\,ms while the rest of its $32$-rank EP group finishes in ${\sim}12$\,ms and idles ${\sim}59$\,ms at the barrier.
The stall bites hardest where the per-sequence CP degree is small: in the tail MBs an EP group spans many CP groups of widely varying sequence lengths, and the bubble ratio reaches $7.4$--$20.0\%$, versus $4$--$7\%$ in the middle MBs and $<4\%$ in the leading ones.
Summed over one iteration, these stalls waste $\sim\!2.3\%$ of aggregate GPU time.

\looseness=-1 \textbf{DP bubble.}
Each EP group is therefore limited by its slowest CP group. 
Unless the CP degree exceeds the EP degree, different DP replicas do not synchronize again until the gradient reduction at the end of the iteration. 
The imbalance does not average out across MBs: a replica assigned a heavier share of sequences remains slower, so the gap between replicas grows throughout the iteration. 
As the first cross-replica barrier, gradient reduction exposes this accumulated gap.
Figure~\ref{fig:dp_imbalance_iters} shows two representative iterations, one severely imbalanced and one mild: the forward/backward boundary is staggered across replicas, so early finishers idle through the gradient reduce-scatter (RS) window until the slowest arrives.
The straggler is large and persistent: the slowest replica's gradient RS runs up to $6.2\times$ as long as the fastest (Figure~\ref{fig:dp_grad_maxmin}), and the RS phase occupies $7\%$ to $27\%$ of the aggregate DP GPU-time (Figure~\ref{fig:dp_grad_ratio}), stretching each iteration by $8$--$12.8\%$.
Prior work has documented DP imbalance under static CP~\cite{ge2025bytescale}.
In DCP, we further show that the DP bubble arises from intra-MB imbalance accumulating across MBs.

\Insight{2}{Balancing MBs is not enough: within one MB, CP groups differ by up to $2.4\times$ in attention computation, and no barrier in the iteration corrects it, so the gap accumulates MB after MB into an $8$--$12.8\%$ DP bubble.
Balancing must reach inside the MB, across CP groups and DP replicas.}

\subsection{Scheduling Overhead}
\label{subsec:overhead}

Beyond load imbalance, Mcore DCP also incurs overhead to compute each schedule.
It runs on the CPU at the head of each iteration's data-fetch path and lies on the critical path: every steady-state iteration opens with a scheduling band spanning all DP replicas (Figure~\ref{fig:dp_imbalance_iters}), during which the GPUs idle.
Across profiled iterations this stall is stable at $8$--$9\%$ of the iteration time.

The cost is not the scheduling decision: choosing a sequence's CP degree and ranks is cheap.
The plan is built incrementally, one sub-sample and one MB at a time. 
For each sub-sample, the scheduler linearly scans the entire HDP pool to choose ranks, rebalance load, and fill idle ranks. 
This cost grows with both the number of sub-samples and the pool width. 
Long-context training makes both large: a batch holds tens of thousands of short sub-samples, while the pool spans tens of thousands of ranks.
What is negligible at small scale thus compounds into the stall we observe.

\Insight{3}{Mcore DCP scans the entire rank pool once per sub-sample, tens of thousands of times per batch, idling GPUs for $8$--$9\%$ of each iteration.
Online planning must preserve placement quality without sweeping the entire pool for every sequence.}

\begin{figure}[h]
    \centering
    \includegraphics[width=0.8\linewidth]{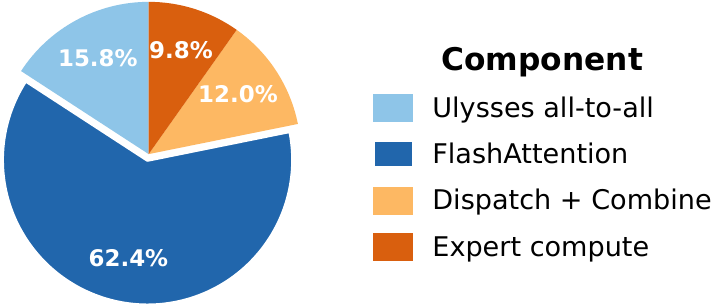}
    \caption{\textit{Forward-pass latency breakdown of one transformer layer in the heaviest-loaded MB.}}
    \label{fig:layer_latency_breakdown}
\end{figure}

\subsection{CP Scaling Dilemma}
\label{subsec:dilemma}

\textbf{The MB that paces the iteration is attention-bound.}
Figure~\ref{fig:layer_latency_breakdown} breaks down one transformer layer on that MB's bottleneck rank. The rank processes a single $256$K-token sequence, to which Mcore's memory-driven policy assigns $\text{cp}{=}32$.
FlashAttention dominates the layer, accounting for $62.4\%$ of its latency and far exceeding either the Ulysses all-to-all or the entire MoE block.
The iteration is therefore paced by the longest sequence's per-rank attention cost, $S^2/\text{cp}$; reducing iteration time requires lowering this term.


\textbf{Balance asks for a larger degree, not a smaller one.}
Every MB uses the same fixed HDP rank pool. 
Reducing the heaviest MB's load therefore requires assigning more ranks to its long sequences, which lowers their $S^2/\text{cp}$ cost.
Mcore packs each sequence onto the fewest ranks that hold it, while ByteScale uses the minimum required number of devices~\cite{nvidia2025dynamiccp, ge2025bytescale}.
Solver-based systems choose group degrees from a finite set of candidate configurations and assign sequences among them~\cite{wang2025flexsp, li2025hydraulis}. 
They can select larger degrees, but still rely on native Ulysses or ring communication and thus inherit the limits analyzed next.
Effective balancing therefore requires both going beyond the memory-minimum degree and executing those larger degrees efficiently; 
the two native CP schemes fail in opposite ways.

\begin{figure}[t]
    \centering
    \includegraphics[width=0.95\linewidth]{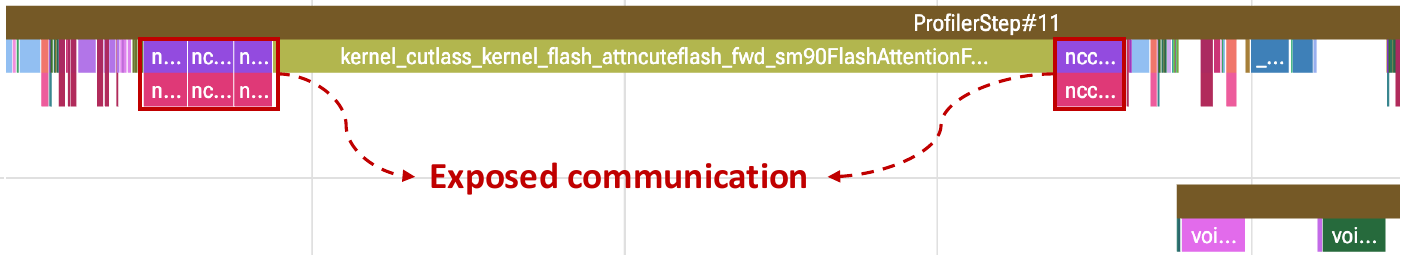}
    \caption{\textit{Ulysses all-to-all (QKV and output transpose) is exposed on the attention critical path.}}
    \label{fig:ulysses_exposed}
\end{figure}

\textbf{Ulysses scales down both computation and communication but stops at the head count.}
Ulysses uses all-to-all transposes to partition attention heads across ranks.
As the degree $P$ grows, both per-rank attention computation and communication fall as $1/P$:
\begin{equation}
T^{\text{uly}}_{\text{comp}} \propto \frac{S^2 h}{P}, \qquad
V^{\text{uly}}_{\text{comm}} \approx \frac{4 S h}{P}.
\label{eq:ulysses}
\end{equation}
This joint reduction is what load balancing requires; otherwise, communication would replace computation as the bottleneck.
First, head partitioning caps $P\!\le\!H/\text{tp}$.
Flagship models typically have at most about $64$ attention heads~\cite{deepseek2026v4, kimi2026k25, glm2026glm5, hy3, hy4preview}, with even fewer full-attention heads in hybrid models~\cite{qwen2026qwen36}; at $\text{tp}{=}2$, Ulysses therefore stops at $\text{cp}{=}32$, short of what the bottleneck MB needs to reach the load target.
Second, the QKV and output all-to-alls remain exposed on the attention critical path, accounting for $4\%$ of end-to-end iteration time in production (Figure~\ref{fig:ulysses_exposed}).

\textbf{Ring reaches any degree, but its traffic never shrinks.}
Ring CP circulates K/V blocks around the CP ring using point-to-point send/recv.
These blocks contain only the $h_{kv}\!\le\!h$ key/value heads.
Unlike Ulysses, ring has no head-count limit and pipelines communication with attention computation.
Its per-rank computation decreases with $P$, but its communication volume does not:
\begin{equation}
T^{\text{ring}}_{\text{comp}} \propto \frac{S^2 h}{P}, \qquad
V^{\text{ring}}_{\text{comm}} \approx 2 S h_{kv}, \qquad
\frac{V^{\text{ring}}_{\text{comm}}}{V^{\text{uly}}_{\text{comm}}} = \frac{P\,h_{kv}}{2h}.
\label{eq:ring}
\end{equation}
Ring traffic is independent of $P$.
It undercuts Ulysses only below $P=2h/h_{kv}$: $\text{cp}{=}16$ for $h{=}64$ and $h_{kv}{=}8$, but $\text{cp}{=}2$ under MLA, where $h_{kv}\!\approx\!h$.
Both crossovers are far below the degrees needed to balance a $256$K sequence.
At those degrees, ring communication sets iteration time; under MLA, it already exceeds attention computation in production and cannot be hidden.

\Insight{4}{\looseness=-1 Balancing the heaviest MB requires increasing cp.
Ulysses shrinks traffic with cp but stops at the head count and leaves its all-to-all exposed.
Ring reaches any cp, but its traffic never shrinks.
A CP scheme must scale beyond the head count without making communication the bottleneck.}

\section{\OURS Overview}
\label{sec:overview}

\subsection{Design Principles}

The four insights of \S\ref{sec:measurement} translate into four principles that guide the design of \OURS.

\looseness=-1 \textbf{(1) Balance computation, not just tokens.}
Insight~1 shows that equal token allocations conceal orders-of-magnitude differences in attention workload, leaving MB execution times $5\times$ apart.
Production DCP must therefore balance attention computation across MBs, not just their tokens, while preserving the memory feasibility it already provides.

\textbf{(2) Balance within and across MBs.}
Insight~2 shows that per-sequence CP leaves CP groups within an MB with widely different computation loads; this skew stalls the MoE all-to-alls and accumulates across MBs into the DP bubble.
Balancing must therefore cover CP groups within each MB and DP replicas across the iteration.

\textbf{(3) Keep online planning lightweight.}
Insight~3 shows that building the plan stalls $8$--$9\%$ of every production iteration, at a cost that grows with both the sequence count and the width of the rank pool.
Planning must therefore avoid whole-pool work as cluster size enlarges both.

\textbf{(4) Scale CP without creating a communication bottleneck.}
Insight~4 shows that balancing the heaviest MB drives the CP degree up rather than down, yet Ulysses and ring communication fail those degrees from opposite directions.
The communication mechanism must therefore reach degrees beyond the attention-head count with traffic that shrinks as the degree grows and can be overlapped.

\subsection{System Overview}

\begin{figure}[t]
    \centering
    \includegraphics[width=\linewidth]{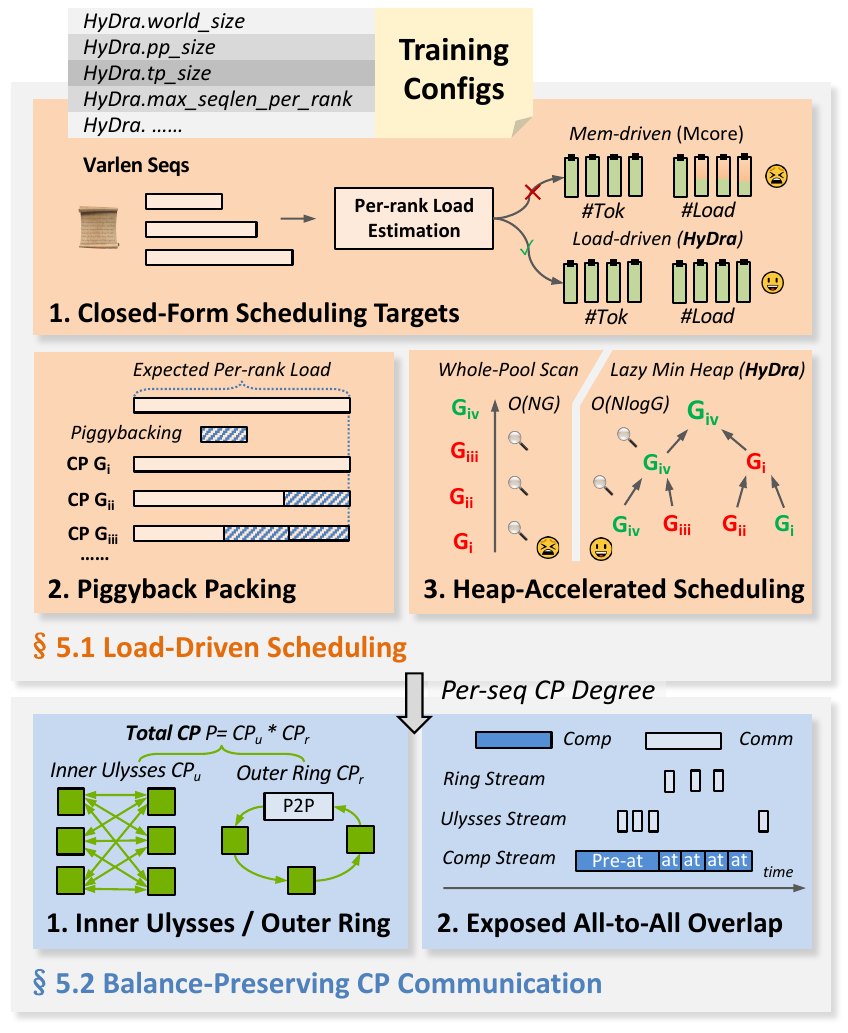}
    \caption{\textit{\OURS overview.}}
    \label{fig:hydra_system_overview}
\end{figure}

\looseness=-1 As shown in Figure~\ref{fig:hydra_system_overview}, \OURS realizes the four principles with two co-designed components, a load-driven scheduler and a balance-preserving CP communication engine, integrated into an existing training stack.

\textbf{(1) Load-driven scheduling.}
At each iteration, the scheduler estimates attention workload from sequence lengths and chooses each sequence's CP degree and placement to balance MBs, their CP groups, and DP replicas against a single per-rank load target, while keeping planning inexpensive.
This component realizes Principles~1--3.

\textbf{(2) Balance-preserving CP communication.}
Reaching that target means raising a heavy sequence's CP degree, which helps only if communication does not become the new critical path.
This component executes the selected degrees beyond native CP limits and hides the exposed transfers behind dependency-free computation, realizing Principle~4.

\looseness=-1 The end-to-end flow is simple: the data path collects sequence lengths, the scheduler produces a placement plan, the data are routed accordingly, and attention executes with the selected CP groups.
Neither component stands alone: the target the scheduler sets is unreachable on native CP, and the degrees the engine executes pay off only when a balanced plan requests them.
Section~\ref{sec:design} presents both designs in detail.

\section{\OURS Design}
\label{sec:design}

\subsection{Load-Driven DCP Scheduling}
\label{subsec:scheduling}

\begin{table}[t]
\centering
\caption{\textit{Notation for \OURS scheduling.}}
\label{tab:notation}
\small
\setlength{\tabcolsep}{4pt}
\renewcommand{\arraystretch}{1.15}
\resizebox{0.9\linewidth}{!}{%
\begin{tabularx}{\linewidth}{@{}l X@{}}
\toprule
\textbf{Symbol} & \textbf{Description} \\
\midrule
\rowcolor{gray!12}\multicolumn{2}{@{}l}{\textit{Workload}}\\
\quad $s,\ s_i,\ s_{\max}$ & sequence length; of sequence $i$; longest in the batch \\
\quad $N$ & \# sequences in the batch \\
\quad $W$ & total quadratic attention work, $\sum_i s_i^2$ \\[1pt]
\rowcolor{gray!12}\multicolumn{2}{@{}l}{\textit{System configuration}}\\
\quad $G$ & \# ranks in the HDP pool (DP$\times$CP) \\
\quad $B$ & per-rank token budget (memory limit) \\
\quad $pp$ & pipeline-parallel depth (\# stages) \\
\quad $h,\ h_{kv}$ & \# query / KV heads per rank after TP ($h_{kv}\!\ll\!h$) \\
\addlinespace[1pt]
\rowcolor{gray!12}\multicolumn{2}{@{}l}{\textit{Scheduler decisions}}\\
\quad $\Lambda,\ \Lambda^{\star}$ & per-rank load target; balance-oriented target \\
\quad $M,\ M^{\star}$ & MB count; target count \\
\quad $C_{\mathrm{mem}}$ & smallest cap fitting the longest sequence within $B$ \\
\quad $C,\ \widehat C^\star,\ C^{\star}$ & CP-degree cap; optimum; power-of-two value \\
\quad $cp(s)$ & CP degree of a length-$s$ sequence \\[1pt]
\rowcolor{gray!12}\multicolumn{2}{@{}l}{\textit{Cap cost model}}\\
\quad $\theta,\ c$ & time per unit per-rank load; fixed cost per MB \\
\quad $\beta,\ \gamma$ & bubble-arm / overhead-arm coefficient \\
\quad $\kappa$ & MBs per unit of cap, $W/(s_{\max}^2G)$ \\[1pt]
\rowcolor{gray!12}\multicolumn{2}{@{}l}{\textit{CP execution}}\\
\quad $P$ & CP degree of a sequence ($=cp(s)$) \\
\quad $cp_u,\ cp_r$ & inner Ulysses / outer ring factor of $P$ \\
\bottomrule
\end{tabularx}%
}
\end{table}

\OURS sizes each sequence's CP degree by its quadratic attention load, not its memory footprint: splitting a length-$s$ sequence over $cp$ ranks yields per-rank load $s^2/cp$, and the heaviest MB paces the iteration (\S\ref{subsec:intermb}).
It then derives closed-form targets, realizes them through piggyback packing, and accelerates planning with heaps.

\subsubsection{Closed-Form Scheduling Targets}

The targets couple per-rank load $\Lambda$ (and thus MB count $M$) with a CP-degree cap $C$; we derive $\Lambda^\star$ and $M^\star$ for fixed $C$, then $C^\star$ and each sequence's $cp(s)$ (Table~\ref{tab:notation}).

\textit{\underline{Balance-oriented load target.}}
Fix the cap $C$ for now.
The target must reconcile two asymmetric limits imposed by long and short sequences.
First, the longest sequence sets a hard floor on the target: even at the maximum degree $C$, the MB holding it carries a per-rank load of at least $s_{\max}^2/C$.
Second, short-sequence MBs can exhaust the per-rank token budget $B$ before reaching a large quadratic-load target.
For a sequence $i$ assigned to $p_i$ ranks, let $x_i=s_i/p_i$ be its rank-local token count.
For the sequences $\mathcal{P}_r$ packed on rank $r$, token occupancy $t_r$ and quadratic attention load $q_r$ satisfy
%
\begin{equation}
\begin{aligned}
t_r &= \sum_{i\in\mathcal P_r} x_i \le B,\\[2pt]
q_r &= \sum_{i\in\mathcal P_r} s_i x_i
     = t_r\bar s_r \le B\bar s_r.
\end{aligned}
\label{eq:packing-density}
\end{equation}
Here $\bar{s}_r=q_r/t_r$ is the token-weighted sequence length: it determines the quadratic work obtained from each packed token, and $t_r\le B$ caps the attainable load at $q_r\le B\bar{s}_r$.
A rank filled with short sequences has a small $\bar{s}_r$, so it exhausts $B$ while $q_r$ remains far below a large $\Lambda$.
Increasing $\Lambda$ beyond this attainable load cannot put more computation into such MBs; 
instead, it leaves them underloaded while allowing long sequences to use fewer CP ranks and approach the higher target, worsening inter-MB skew.

\OURS therefore sets the load target to this floor,
\begin{equation}
\Lambda^\star=\frac{s_{\max}^2}{C}.
\label{eq:load-target}
\end{equation}
A lower target is infeasible for the capped longest sequence, while a higher one is harder for token-limited short-sequence MBs to attain, so $\Lambda^\star$ is at once the smallest feasible target and the easiest to approach.
It is selected primarily for MB balance, rather than as the unconstrained minimizer of an idealized latency function.

\textit{\underline{Target MB count.}}
Let $\bar q$ be the realized average per-rank load.
Work conservation gives $W=MG\bar q$.
Under balanced packing, $\bar q\approx\Lambda^\star$, yielding the target
%
\begin{equation}
M^\star
=\left\lceil\frac{W}{G\Lambda^\star}\right\rceil
=\left\lceil\frac{CW}{G s_{\max}^{2}}\right\rceil.
\label{eq:mstar}
\end{equation}
$M^\star$ is a planning target rather than the realized count.
Token limits, power-of-two CP groups, and group alignment may push packing off it; PP-bubble and virtual-pipeline constraints then enforce feasibility.

\textit{\underline{Closed-form CP cap.}}
The cap is the only value left to choose, and choosing it is a trade-off.
Since $\Lambda^\star=s_{\max}^2/C$ and $M^\star\approx\kappa C$ with $\kappa=W/(s_{\max}^2G)$, a larger cap makes every MB smaller and the pipeline bubble shorter, but it also creates more MBs, and each MB carries a fixed cost $c$,
%
\begin{equation}
\begin{aligned}
T(C)
&\approx \frac{\theta W}{G}
       +(pp-1)\theta\Lambda^\star+cM^\star\\[3pt]
&\approx \underbrace{T_0}_{\text{compute}}
       +\underbrace{\beta/C}_{\text{bubble}}
       +\underbrace{\gamma C}_{\text{overhead}}.
\end{aligned}
\label{eq:capcost}
\end{equation}
Without balance, iteration time is a graph of waits, each a maximum over ranks, and no closed form exists.
Under a perfectly balanced schedule the equation is exact, since every MB takes the same time $\theta\Lambda^\star$, 1F1B takes $M+pp-1$ MB times, and work conservation turns the first $M$ of them into $\theta W/G$.
The targets above drive the schedule toward that balance, so we use it as an approximation.
The coefficients in Eq.~\eqref{eq:capcost} are
\begin{equation}
T_0=\theta W/G,\quad
\beta=(pp-1)\,\theta\,s_{\max}^2,\quad
\gamma=c\,\kappa .
\label{eq:bgamma}
\end{equation}
$T_0$ is the total work divided over the ranks, which the cap does not change, so only the last two terms decide the best cap.
The $\beta/C$ term decreases as the cap grows while $\gamma C$ rises, so the curve is convex and has one lowest point,
%
\begin{equation}
\begin{gathered}
T'(C)=0
\quad\Longleftrightarrow\quad
C^2=\frac{\beta}{\gamma},\qquad C>0,\\[5pt]
\widehat C^\star
=\sqrt{\frac{\beta}{\gamma}}
=s_{\max}^{2}\sqrt{\frac{(pp-1)\theta G}{cW}}
\end{gathered}
\label{eq:cstar-cont}
\end{equation}
Only $\theta/c$ has to be measured, and a ratio of two costs is easier to measure reliably than either cost by itself.
Memory and power-of-two groups then decide the cap we can run.
Let $\lceil x\rceil_2$ denote the smallest value in $\{1,2,4,\ldots\}$ that is at least $x$; then
%
\begin{equation}
\begin{aligned}
C_{\mathrm{mem}}
&=\left\lceil\frac{s_{\max}}{B}\right\rceil_2,\\[3pt]
C^\star
&=\min\!\left\{G,\,
  \Bigl\lceil\max\{C_{\mathrm{mem}},\widehat C^\star\}\Bigr\rceil_2
  \right\}.
\end{aligned}
\label{eq:cstar}
\end{equation}
We round up rather than down, because a larger cap only lowers the target, and MBs made of short sequences can always reach a lower target.
Rounding to powers of two also means that a small error in $\theta/c$ usually gives the same cap.

\textit{\underline{Load-driven CP degree.}}
A length-$s$ sequence at degree $cp$ also places $s/cp$ tokens per rank, so meeting the computation target requires $cp\ge\lceil s^2/\Lambda^\star\rceil$, while fitting within the token budget requires $cp\ge\lceil s/B\rceil$.
Megatron-Core enforces only the latter and thus leaves the computation skew measured in Section~\ref{sec:measurement}.
\OURS uses the smallest power-of-two degree satisfying both:
%
\begin{equation}
\operatorname{cp}(s)
=\min\!\left\{C^\star,\,
  \left\lceil\max\!\left\{
    \frac{s^2}{\Lambda^\star},\frac{s}{B}
  \right\}\right\rceil_2
  \right\}.
\label{eq:cp}
\end{equation}
%
Because $\Lambda^\star=s_{\max}^2/C^\star$ and $C^\star\ge C_{\mathrm{mem}}$, both per-sequence requirements fit within the cap, so the outer minimum preserves them.
An open group $g$ is feasible for sequence $s$ if adding it keeps both $\mathrm{load}(g)+s^2/|g|\le\Lambda^\star$ and $\mathrm{tokens}(g)+s/|g|\le B$.
\begin{algorithm}[t]
    \DontPrintSemicolon
    \SetAlgoNlRelativeSize{-1}
    \SetNlSty{textbf}{}{}
    \SetKwInOut{Input}{Input}\SetKwInOut{Output}{Output}
    \newcommand{\algcommentfont}[1]{{\itshape\textcolor{black!55}{#1}}}
    \SetKwComment{tcc}{\textcolor{RoyalBlue}{$\triangleright$}\ }{}
    \SetKwComment{note}{}{}
    \SetCommentSty{algcommentfont}
    \SetKwFunction{Cp}{cp}
    \SetKw{KwAnd}{and}
    \SetKw{KwBreak}{break}
    \caption{\OURS DCP Scheduling}
    \label{alg:hydra-scheduling}
    \Input{documents $\{s_i\}$ and $s_{\max}$; rank pool $G$; cap $C\!=\!C^{\star}$ (Eq.~\eqref{eq:cstar}); budget $B$}
    \Output{$\mathcal{G}[m][r]$: doc. lists for MB $m$ and rank $r$}
    \BlankLine
    
    \tcc{\textcolor{RoyalBlue}{Phase 1: closed-form load targets}}
    $W \leftarrow \sum_i s_i^2$;\quad $\Lambda^{\star} \leftarrow s_{\max}^2/C$\note*[r]{target per-rank load}
    $M^{\star} \leftarrow \big\lceil C\,W / (s_{\max}^2 G) \big\rceil$\note*[r]{Eq.~\eqref{eq:mstar}}
    \BlankLine
    
    \tcc{\textcolor{RoyalBlue}{Phase 2: longest-first piggyback packing}}
    sort $\{s_i\}$ descending;\quad $m \leftarrow 0$\;
    \While{documents remain}{
      open MB $m$ over all $G$ ranks\;
      \While{documents remain}{
        $s \leftarrow$ next document;\quad $k \leftarrow \Cp{s}$\note*[r]{Eq.~\eqref{eq:cp}}
        \uIf{free ranks $\ge k$}{
          $g \leftarrow$ new $k$-aligned group on free ranks 
        }
        \uElseIf{$\exists$ feasible group $g$ with $|g|\!\ge\!k$}{
          $g \leftarrow$ smallest such group\;
          break ties by load\note*[r]{piggyback}
        }
        \lElse{\KwBreak}
        pack $s$ into $g$;\ update load and occupancy\;
        \lIf{$k$ has shrunk \KwAnd\ MB balanced}{\KwBreak}
      }
      backfill idle ranks;\quad $m \leftarrow m+1$\;
    }
    \BlankLine
    
    \tcc{\textcolor{RoyalBlue}{Phase 3: finalize}}
    broadcast $m$;\quad \Return $\mathcal{G}$\;
\end{algorithm}

\subsubsection{Piggyback Packing}
Algorithm~\ref{alg:hydra-scheduling} places sequences longest-first, one MB at a time across all $G$ ranks.
For a sequence needing degree $k$, it first opens a new $k$-aligned group if enough free ranks are available.
Otherwise, it \emph{piggybacks} on the smallest feasible open group of size at least $k$, breaking ties by load.
Alignment keeps every group within a prebuilt power-of-two rank chunk and preserves ring ordering.
This packs short sequences into the gaps left by wide long-sequence groups, cutting both idle ranks and the MB count.
After closing each MB, \OURS backfills idle ranks; once all sequences are placed, it broadcasts the realized MB count so every pipeline stage uses the same count.
Because every MB is packed across the whole HDP pool against one target $\Lambda^\star$, the balance extends past the pipeline to the CP and EP groups inside an MB and the DP replicas that synchronize after it (Insight~2).

\subsubsection{Heap-Accelerated Scheduling}
A direct implementation scans every compatible group per sequence and re-sweeps the rank pool on each backfill, recreating the whole-pool traversals of Insight~3.
\OURS partitions open groups by CP degree and keeps, per class, a lazy min-heap keyed by group load and a companion max-heap keyed by remaining token capacity.
Stale load-heap entries are discarded on access, and a candidate that violates the token budget is \emph{demoted} to the capacity heap rather than rescanned.
Re-admission is cheap because a class fixes the divisor $|g|$ and sequences arrive longest-first, so the per-rank demand a class sees never grows: each group is touched $O(1)$ times per placement instead of once per search.
If the load-heap minimum already reaches $\Lambda^\star$, the class is pruned outright, since loads only grow and no later sequence can enter it.
A third min-heap backfills idle ranks by repeatedly doubling the smallest group, and the pool maximum read by the MB-balance test is maintained incrementally, as every rank in a CP group carries the same load.
These structures reproduce the linear search's least-loaded placement and deterministic tie-breaking exactly, while replacing its $O(NG)$ group and rank sweeps with $O(N\log G)$.
\OURS lowers this cost rather than hiding it behind training, since planning and kernel launch share one CPU.

\subsection{Balance-Preserving CP Communication}
\label{subsec:comm}

Balancing raises a heavy sequence's CP degree $P$, reducing its per-rank load to $s^2/P$.
Native CP cannot exploit this: Ulysses is head-count limited and exposed, while ring has a degree-independent traffic floor (\S\ref{subsec:dilemma}).
\OURS combines Ulysses--ring factorization with exposed all-to-all overlap.

\subsubsection{Inner Ulysses $\times$ Outer Ring}

The factorization runs each selected degree on an inner Ulysses dimension inside a shallow outer ring~\cite{fang2024usp}.

\textit{\underline{Beyond the head-count limit.}}
We factor the scheduler-selected degree into an inner Ulysses group over heads and an outer ring over the sequence:
\begin{equation}
P=cp_u\cdot cp_r,\qquad cp_u=\min(P,h),\quad cp_r=P/cp_u .
\label{eq:usp}
\end{equation}
For $P\le h$ this is pure Ulysses; beyond $h$ the head dimension saturates and the outer ring supplies the residual factor, so the scheduler may request any $P$ up to $C^\star$ while per-rank computation stays $s^2/P$.

\vspace{0.5em}
\begin{figure}[b]
    \centering
    \includegraphics[width=\linewidth]{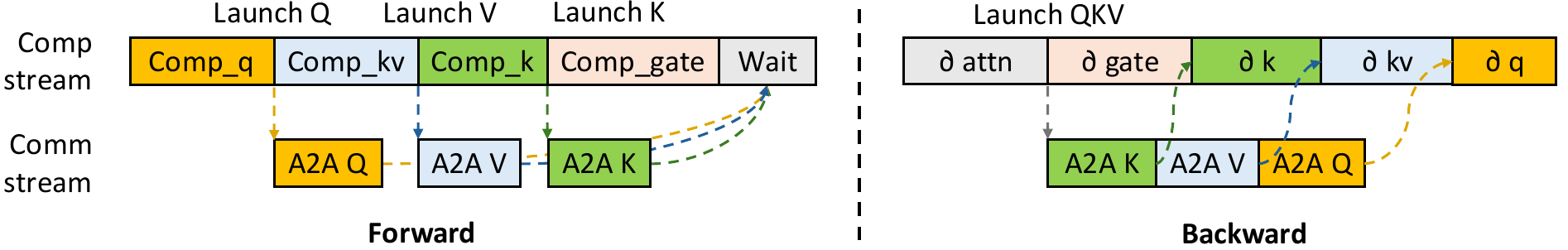}
    \caption{\textit{Asynchronous all-to-all schedule in gated MLA, forward and backward.}}
    \label{fig:async_a2a}
\end{figure}

\begin{figure*}[t]
    \centering
    \begin{subfigure}[b]{0.30\textwidth}
        \centering
        \includegraphics[width=\linewidth]{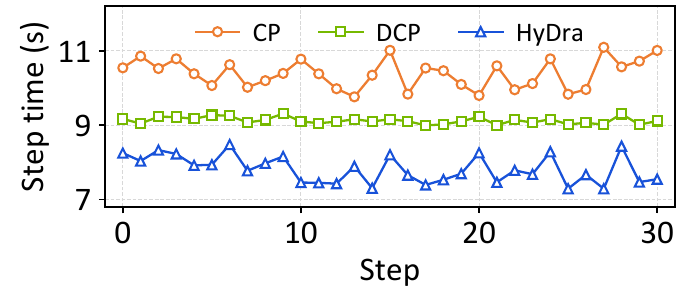}
        \subcaption{\textit{32K, per step}}
        \label{fig:512_h20_series_32k}
    \end{subfigure}
    \hfill
    \begin{subfigure}[b]{0.175\textwidth}
        \centering
        \includegraphics[width=\linewidth]{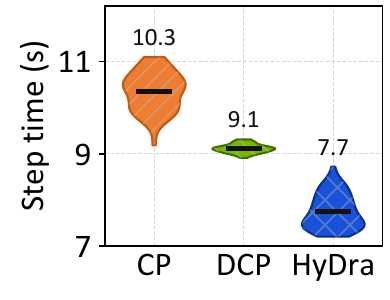}
        \subcaption{\textit{32K, range}}
        \label{fig:512_h20_range_32k}
    \end{subfigure}
    \hfill
    \begin{subfigure}[b]{0.30\textwidth}
        \centering
        \includegraphics[width=\linewidth]{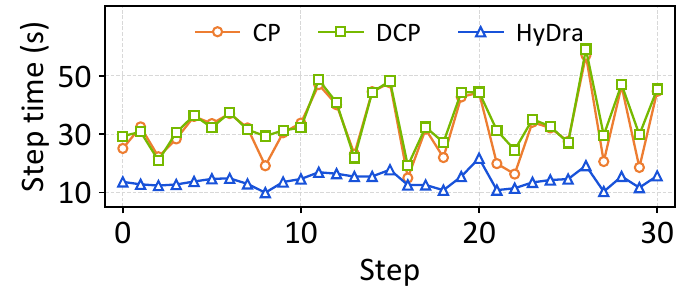}
        \subcaption{\textit{256K, per step}}
        \label{fig:512_h20_series_256k}
    \end{subfigure}
    \hfill
    \begin{subfigure}[b]{0.175\textwidth}
        \centering
        \includegraphics[width=\linewidth]{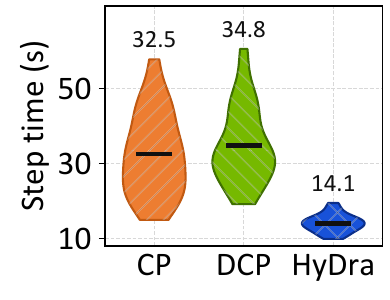}
        \subcaption{\textit{256K, range}}
        \label{fig:512_h20_range_256k}
    \end{subfigure}
    \caption{\textit{Step time on 512 H20 GPUs.}}
    \label{fig:512_h20_step_time}
\end{figure*}

\begin{figure*}[t]
    \centering
    \begin{minipage}[b]{0.325\textwidth}
        \setcounter{subfigure}{0}
        \centering
        \begin{subfigure}[b]{0.47\linewidth}
            \centering
            \includegraphics[width=\linewidth]{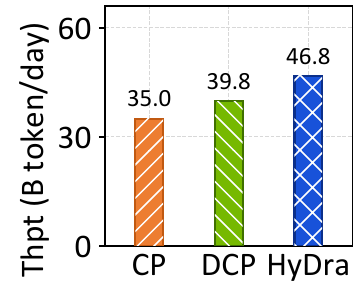}
            \subcaption{\textit{32K}}
            \label{fig:512_h20_thpt_32k}
        \end{subfigure}
        \hfill
        \begin{subfigure}[b]{0.47\linewidth}
            \centering
            \includegraphics[width=\linewidth]{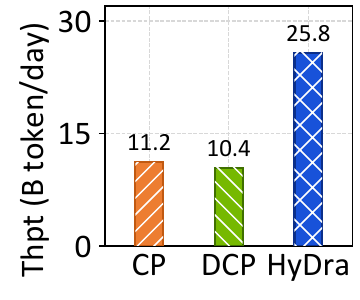}
            \subcaption{\textit{256K}}
            \label{fig:512_h20_thpt_256k}
        \end{subfigure}
        \caption{\textit{Training throughput.}}
        \label{fig:512_h20_throughput}
    \end{minipage}
    \hfill
    \begin{minipage}[b]{0.655\textwidth}
        \setcounter{subfigure}{0}
        \centering
        \begin{subfigure}[b]{0.235\linewidth}
            \centering
            \includegraphics[width=\linewidth]{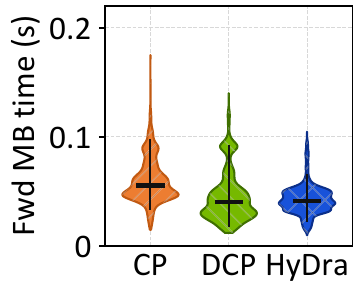}
            \subcaption{\textit{Fwd, 32K}}
            \label{fig:512_h20_mb_fwd_32k}
        \end{subfigure}
        \hfill
        \begin{subfigure}[b]{0.235\linewidth}
            \centering
            \includegraphics[width=\linewidth]{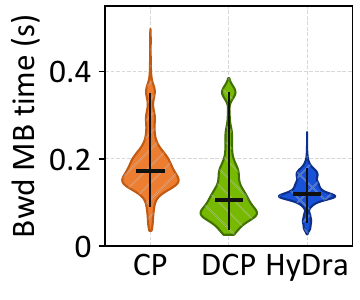}
            \subcaption{\textit{Bwd, 32K}}
            \label{fig:512_h20_mb_bwd_32k}
        \end{subfigure}
        \hfill
        \begin{subfigure}[b]{0.235\linewidth}
            \centering
            \includegraphics[width=\linewidth]{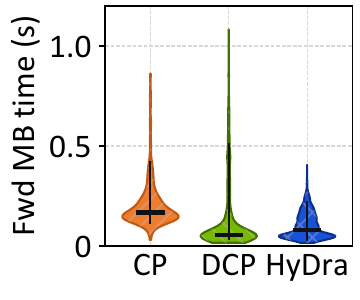}
            \subcaption{\textit{Fwd, 256K}}
            \label{fig:512_h20_mb_fwd_256k}
        \end{subfigure}
        \hfill
        \begin{subfigure}[b]{0.235\linewidth}
            \centering
            \includegraphics[width=\linewidth]{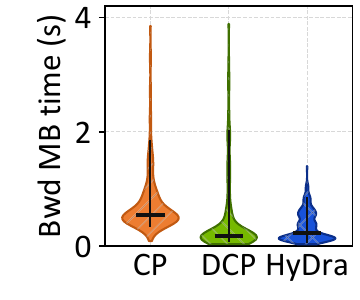}
            \subcaption{\textit{Bwd, 256K}}
            \label{fig:512_h20_mb_bwd_256k}
        \end{subfigure}
        \caption{\textit{Per-microbatch forward and backward time on 512 H20 GPUs.}}
        \label{fig:512_h20_mb_time}
    \end{minipage}
\end{figure*}

\textit{\underline{Keeping the outer ring shallow.}}
Per-rank traffic splits across the two dimensions:
\begin{equation}
V(P)\;\approx\;\underbrace{\Theta \big(s\,h/P\big)}_{\text{inner all-to-all}}
\;+\;\underbrace{\Theta \big(s\,h_{kv}\,\tfrac{cp_r-1}{cp_r}\big)}_{\text{outer ring}} .
\label{eq:comm}
\end{equation}
The Ulysses term falls as $1/P$ while the ring term flattens as $cp_r$ grows; the decomposition does not remove that floor but confines it to the residual factor $cp_r=P/h$ left after the Ulysses dimension saturates.
At $P=256$ and $h=64$ the outer ring spans only four ranks, and its payload is KV-only: small under GQA, though under MLA the up-projected K/V restores a full per-head payload ($h_{kv}\!\approx\!h$).

\subsubsection{Exposed All-to-All Overlap}

The inner all-to-all still lies on the attention critical path, so instead of materializing the query, key, and value and then issuing three blocking collectives, \OURS dispatches each all-to-all on a separate communication stream as soon as its tensor reaches final form (Figure~\ref{fig:async_a2a}).
The mutually independent streams of the gated MLA~\cite{qiu2025gatedattentionlargelanguage} projection fix the launch order: the query goes first, right after its RoPE, covered by the KV down- and up-projections; the value is final once the up-projection output is split, covered by the RoPE and concatenation the key still needs; the key goes last, covered by the gate projection, which needs no context-parallel communication at all.
Deferring all three waits until after the gate maximizes every overlap window.
Backward needs no extra scheduling: launch and wait are paired autograd functions that share one communication handle and swap roles under differentiation, so the adjoint of the compute hiding a collective lands exactly between the two nodes and hides the corresponding gradient collective.
The shallow outer ring pipelines with attention as in standard ring CP.
Together, the factorization and the overlap let a larger $P$ deliver its $s^2/P$ latency reduction instead of a new communication tail.

\begin{figure}[t]
    \centering
    \captionsetup[subfigure]{skip=-0.5pt}
    \includegraphics[width=0.9\linewidth]{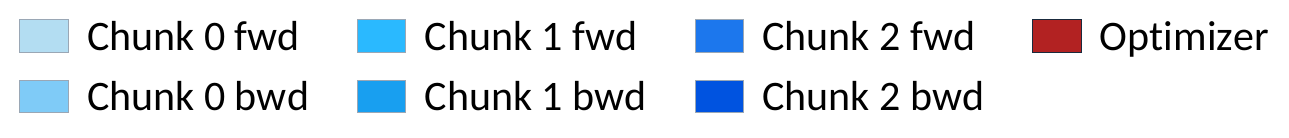}\par
    \vspace{2pt}
    \begin{subfigure}[t]{\linewidth}
        \raggedright
        \includegraphics[height=0.2\linewidth]{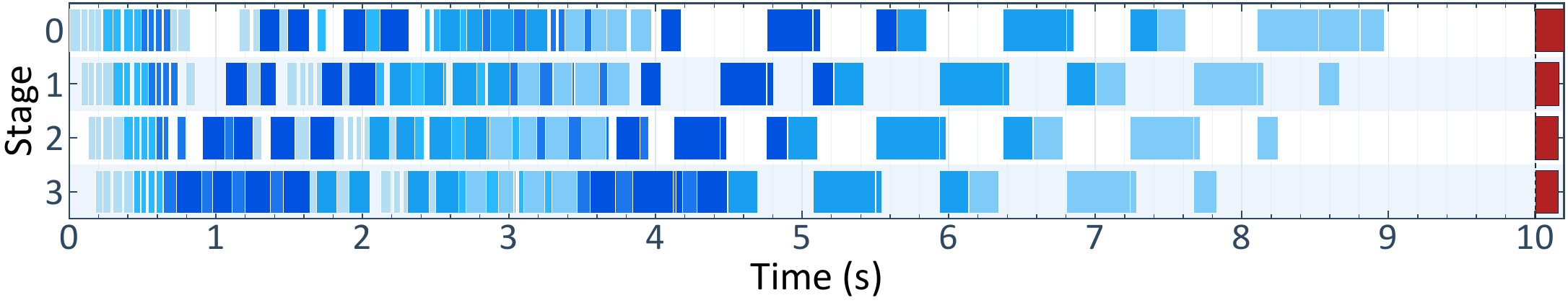}
        \vspace{-10pt}
        \subcaption{\textit{Static CP, 32K}}
        \label{fig:512_pp_cp_32k}
    \end{subfigure}
    \begin{subfigure}[t]{\linewidth}
        \raggedright
        \includegraphics[height=0.2\linewidth]{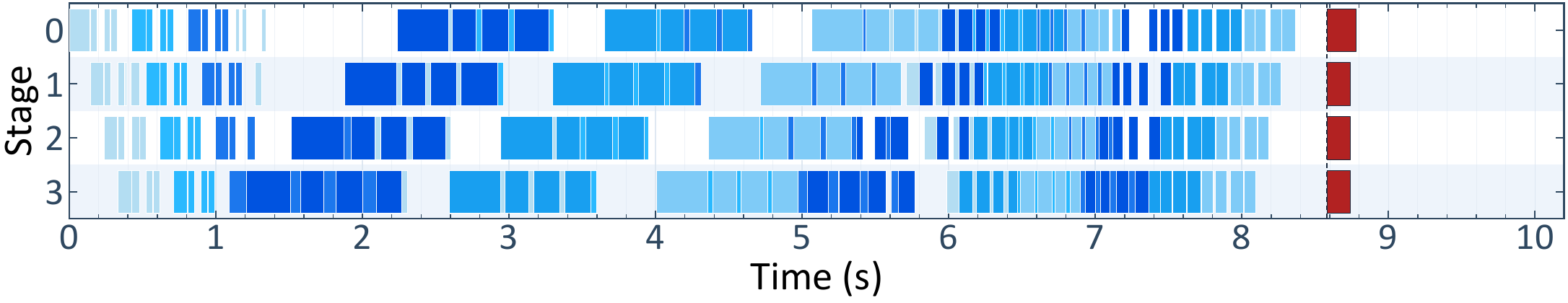}
        \vspace{-10pt}
        \subcaption{\textit{Mcore DCP, 32K}}
        \label{fig:512_pp_dcp_32k}
    \end{subfigure}
    \begin{subfigure}[t]{\linewidth}
        \raggedright
        \includegraphics[height=0.2\linewidth]{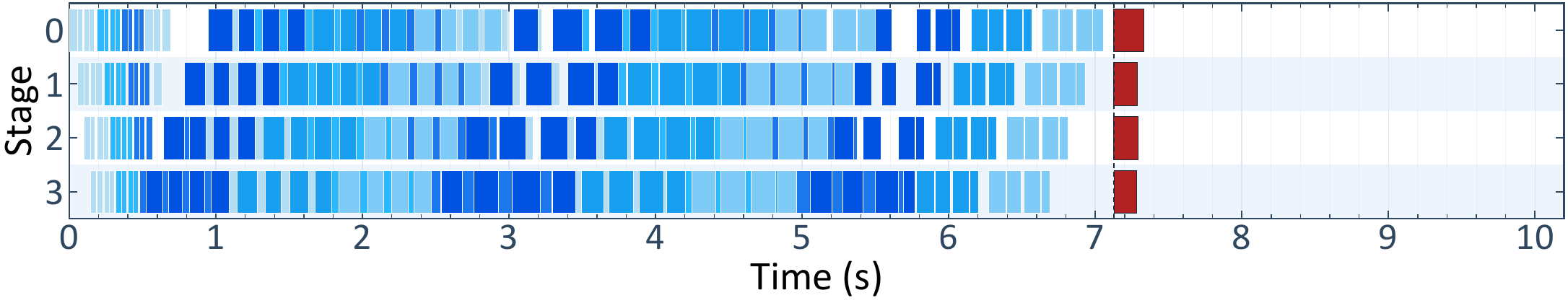}
        \vspace{-10pt}
        \subcaption{\textit{\OURS, 32K}}
        \label{fig:512_pp_hydra_32k}
    \end{subfigure}
    \begin{subfigure}[t]{\linewidth}
        \raggedright
        \includegraphics[height=0.2\linewidth]{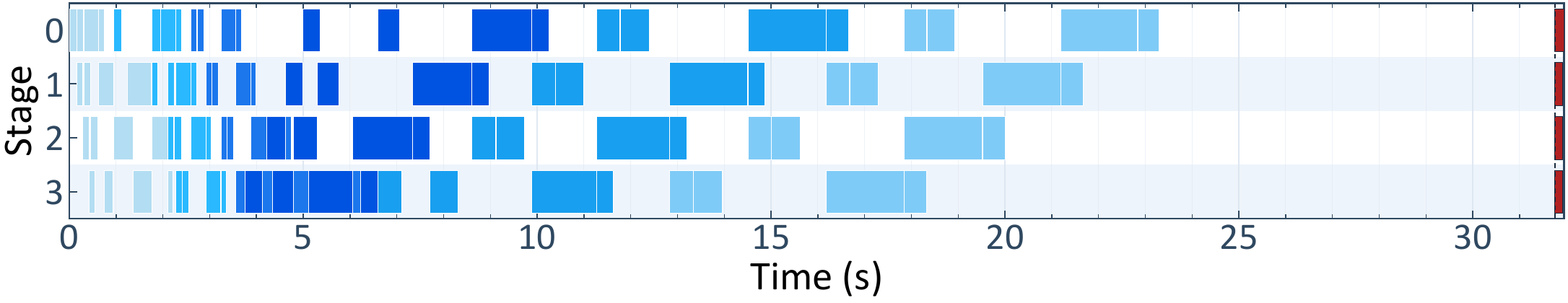}
        \vspace{-10pt}
        \subcaption{\textit{Static CP, 256K}}
        \label{fig:512_pp_cp_256k}
    \end{subfigure}
    \begin{subfigure}[t]{\linewidth}
        \raggedright
        \includegraphics[height=0.2\linewidth]{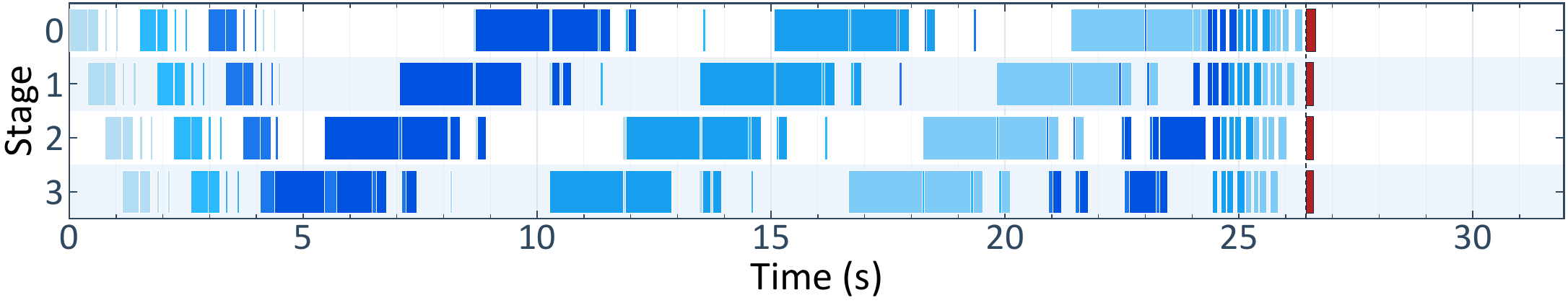}
        \vspace{-10pt}
        \subcaption{\textit{Mcore DCP, 256K}}
        \label{fig:512_pp_dcp_256k}
    \end{subfigure}
    \begin{subfigure}[t]{\linewidth}
        \raggedright
        \includegraphics[height=0.2\linewidth]{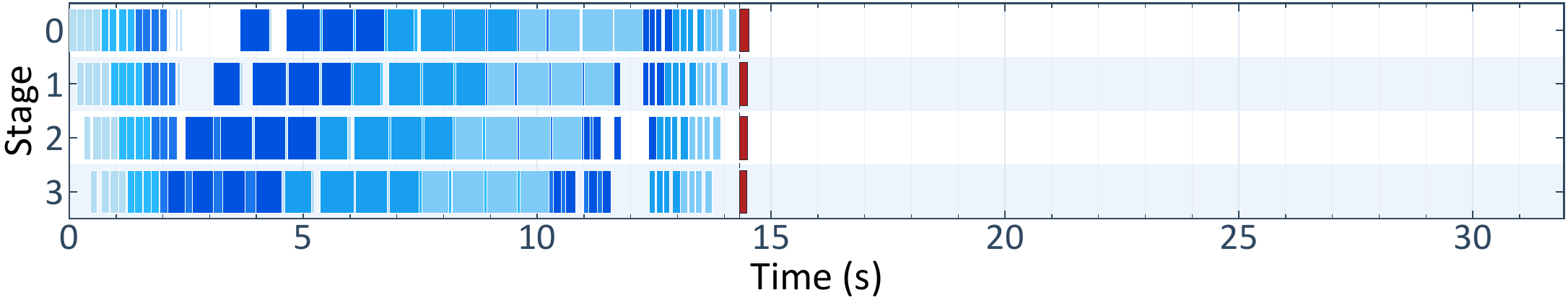}
        \vspace{-10pt}
        \subcaption{\textit{\OURS, 256K}}
        \label{fig:512_pp_hydra_256k}
    \end{subfigure}
    \caption{\textit{One iteration timeline across the four PP stages on 512 H20 GPUs.}}
    \label{fig:512_h20_pp}
\end{figure}

\begin{figure}[t]
    \centering
    \captionsetup[subfigure]{skip=-0.5pt}
    \begin{subfigure}[t]{\linewidth}
        \raggedright
        \includegraphics[width=\linewidth]{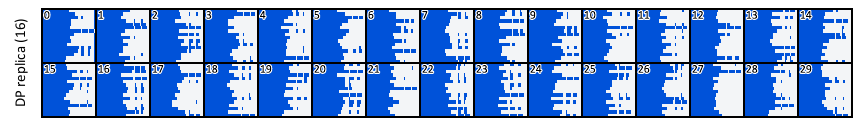}
        \vspace{-10pt}
        \subcaption{\textit{Static CP, 32K}}
        \label{fig:512_dp_cp_32k}
    \end{subfigure}
    \begin{subfigure}[t]{\linewidth}
        \raggedright
        \includegraphics[width=\linewidth]{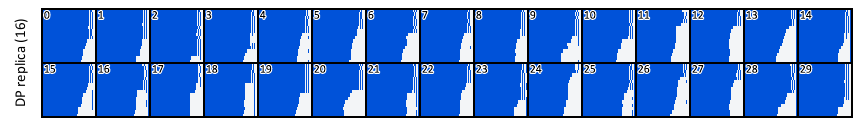}
        \vspace{-10pt}
        \subcaption{\textit{Mcore DCP, 32K}}
        \label{fig:512_dp_dcp_32k}
    \end{subfigure}
    \begin{subfigure}[t]{\linewidth}
        \raggedright
        \includegraphics[width=\linewidth]{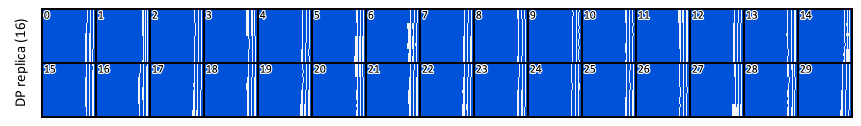}
        \vspace{-10pt}
        \subcaption{\textit{\OURS, 32K}}
        \label{fig:512_dp_hydra_32k}
    \end{subfigure}
    \begin{subfigure}[t]{\linewidth}
        \raggedright
        \includegraphics[width=\linewidth]{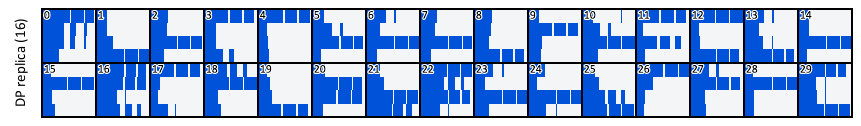}
        \vspace{-10pt}
        \subcaption{\textit{Static CP, 256K}}
        \label{fig:512_dp_cp_256k}
    \end{subfigure}
    \begin{subfigure}[t]{\linewidth}
        \raggedright
        \includegraphics[width=\linewidth]{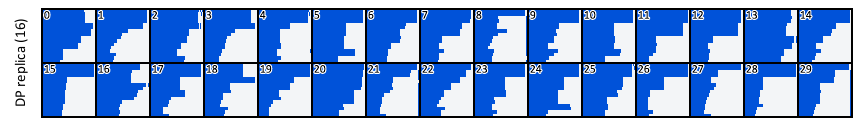}
        \vspace{-10pt}
        \subcaption{\textit{Mcore DCP, 256K}}
        \label{fig:512_dp_dcp_256k}
    \end{subfigure}
    \begin{subfigure}[t]{\linewidth}
        \raggedright
        \includegraphics[width=\linewidth]{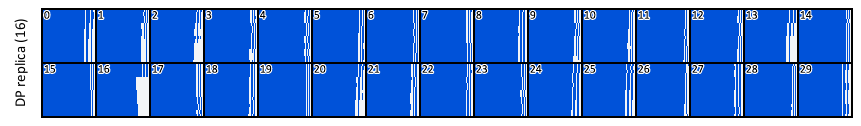}
        \vspace{-10pt}
        \subcaption{\textit{\OURS, 256K}}
        \label{fig:512_dp_hydra_256k}
    \end{subfigure}
    \caption{\textit{Gradient reduce-scatter (gray) of the 16 DP replicas, one panel per step, on 512 H20 GPUs.}}
    \label{fig:512_h20_dp}
\end{figure}

\section{Evaluation}
\label{sec:evaluation}


\subsection{Experimental Setup}
\label{subsec:setup}

\textbf{Clusters.}
We evaluate \OURS in three settings of increasing scale: a \textbf{512}-GPU NVIDIA H20 testbed (\S\ref{subsec:testbed}), a production job on \textbf{2,048} NVIDIA H800 GPUs (\S\ref{subsec:production}), and simulation at 8K--40K GPUs (\S\ref{subsec:simulation}).
Both clusters place 8 GPUs per NVLink node and connect nodes over RoCE.

\textbf{Model.}
Both hardware settings train the same internal MLA-based model\footnote{Model details are withheld for confidentiality.}.

\textbf{Dataset.}
We use two long-context training datasets, one at a 32K context length and one at 256K, whose sequence lengths span orders of magnitude.

\textbf{Parallelism.}
Both settings use $tp1,pp4,ep8$, with a 4K-token per-rank budget at 32K context and 8K at 256K.

\textbf{Baselines.}
On hardware we compare against two production baselines: static CP at the smallest degree that fits the longest sequence, $cp{=}8$ at 32K and $cp{=}32$ at 256K, and Mcore DCP~\cite{nvidia2025dynamiccp}, which sizes CP per sequence.
In simulation we additionally compare against a broader set of baselines~\footnote{These baselines are either closed-source or never validated at production scale, so we can only compare against them in simulation.}, including ByteScale~\cite{ge2025bytescale}, WLB-LLM~\cite{wang2025wlbllm}, and FlexSP~\cite{wang2025flexsp}.

\subsection{Testbed Performance}
\label{subsec:testbed}

\textbf{Throughput.}
On the 512-GPU testbed, \OURS is the fastest system at both context lengths, and its margin widens as the context grows.
At 32K, it raises end-to-end training throughput by $1.19$--$1.52\times$ (averaging $1.34\times$) over static CP and by $1.09$--$1.25\times$ (averaging $1.18\times$) over Mcore DCP, sustaining $46.8$ B tokens per day against $35.0$ B and $39.8$ B, and cutting the mean step time to $7.7$\,s from $10.3$\,s and $9.1$\,s (Figures~\ref{fig:512_h20_range_32k} and~\ref{fig:512_h20_thpt_32k}).
At 256K, the margins grow to $1.19$--$3.00\times$ (averaging $2.30\times$) over static CP and $1.42$--$3.08\times$ (averaging $2.48\times$) over Mcore DCP: $25.8$ B tokens per day against $11.2$ B and $10.4$ B, at a mean step time of $14.1$\,s against $32.5$\,s and $34.8$\,s (Figures~\ref{fig:512_h20_range_256k} and~\ref{fig:512_h20_thpt_256k}).
Mcore DCP is the stronger baseline at 32K but falls behind static CP at 256K, where, as \S\ref{sec:measurement} shows, its memory-driven degrees leave the heavy MBs unbalanced.
The per-step series make the gap visible: \OURS holds a nearly flat step time, while both baselines fluctuate from step to step, most severely at 256K (Figures~\ref{fig:512_h20_series_32k} and~\ref{fig:512_h20_series_256k}).
The same imbalance appears one level down, on individual MBs: both baselines spread MB times widely and trail long tails of slow MBs, whereas \OURS compresses the distribution at both context lengths and in both directions (Figure~\ref{fig:512_h20_mb_time}).

\textbf{PP bubble.}
Tracing where each iteration goes explains the gap.
At 32K, \OURS leaves the smallest pipeline bubble, averaging $17.5\%$ of the iteration and staying within $13.9$--$20.9\%$, against $24.7\%$ ($13.6$--$35.5\%$) under static CP and $26.9\%$ ($21.7$--$31.6\%$) under Mcore DCP.
At 256K the baselines diverge sharply: static CP averages $36.5\%$ ($10.1$--$64.7\%$) and Mcore DCP $55.7\%$ ($42.0$--$65.5\%$), since a memory-driven degree keeps the heaviest MB heavy and every stage waits on it once per microbatch.
\OURS averages $23.3\%$ ($18.0$--$34.4\%$), the only system whose pipeline idle stays both low and narrow as the context grows.
Figure~\ref{fig:512_h20_pp} shows this on the pipeline: under Mcore DCP the four stages are pocked with white gaps that recur through the step and widen at 256K (Figures~\ref{fig:512_pp_dcp_32k} and~\ref{fig:512_pp_dcp_256k}), whereas under \OURS the stages stay packed and the iteration ends far earlier (Figures~\ref{fig:512_pp_hydra_32k} and~\ref{fig:512_pp_hydra_256k}).

\begin{figure}[t]
    \centering
    \includegraphics[width=0.95\linewidth]{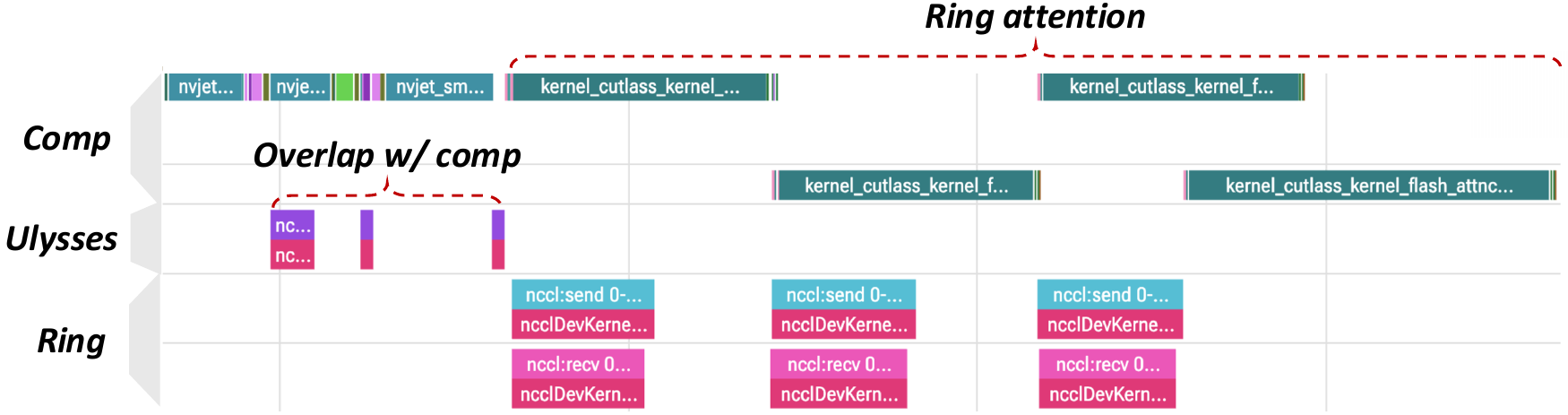}
    \caption{\textit{Attention trace in \OURS.}}
    \label{fig:async_ulysses}
\end{figure}

\begin{figure}[t]
    \centering
    \captionsetup{skip=-3pt}
    \begin{minipage}[b]{0.48\linewidth}
        \centering
        \includegraphics[width=\linewidth]{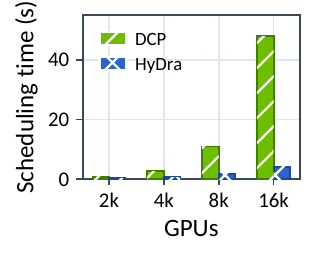}
        \caption{\textit{Scheduling time.}}
        \label{fig:scheduling_overhead}
        \Description{A bar chart compares the scheduling time of Mcore DCP and HyDra from 2K to 16K GPUs.}
    \end{minipage}
    \hfill
    \begin{minipage}[b]{0.48\linewidth}
        \centering
        \includegraphics[width=\linewidth]{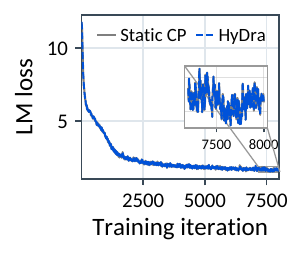}
        \caption{\textit{Training loss.}}
        \label{fig:loss}
        \Description{A line chart compares the language-modeling loss curves of HyDra and static CP.}
    \end{minipage}
\end{figure}

\textbf{DP bubble.}
Static CP suffers the largest DP bubble, averaging $15.3\%$ ($0.5$--$33.3\%$) of the iteration at 32K and $24.5\%$ ($0.1$--$67.4\%$) at 256K, because a fixed degree never rebalances computation across replicas.
In Figures~\ref{fig:512_pp_cp_32k} and~\ref{fig:512_pp_cp_256k} the bubble is the wide empty span between the last chunk's backward kernels and the optimizer, where rank 0 has finished its share and waits for the slowest replica; under \OURS the optimizer follows the last backward almost immediately.
Figure~\ref{fig:512_h20_dp} shows the same effect across all 16 replicas: the gray bands are the reduce-scatter itself, and since a replica stays in the collective until the slowest one arrives, wider bands mean more DP idle.
The collective takes $34.8\%$ of the iteration on average under static CP at 32K and $39.3\%$ at 256K, against $4.2\%$ under Mcore DCP at both lengths and $5.3\%$ and $4.3\%$ under \OURS, whose bands stay thin and aligned.
Sizing CP per sequence largely removes this, but not reliably: Mcore DCP averages $0.9\%$ ($0.6$--$2.4\%$) at 32K and $2.1\%$ ($0.1$--$20.9\%$) at 256K, where its worst steps still idle over a fifth of the iteration.
\OURS is the only system that stays low at both context lengths, averaging $1.0\%$ ($0.7$--$2.9\%$) at 32K and $0.8\%$ ($0.2$--$2.7\%$) at 256K.
The two effects compound: the compute stream is busy $81.5\%$ of the iteration under \OURS at 32K, against $60.0\%$ under static CP and $72.2\%$ under Mcore DCP, and $76.0\%$ at 256K, against $39.0\%$ and $42.2\%$.

\begin{figure}[t]
    \centering
    \begin{subfigure}[t]{0.32\linewidth}
        \centering
        \includegraphics[width=\linewidth]{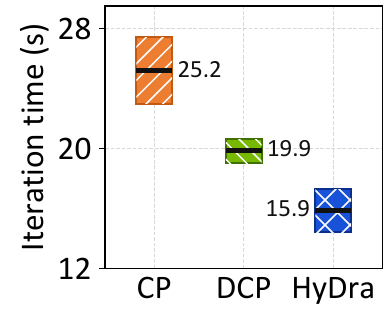}
        \subcaption{\textit{Iteration time}}
        \label{fig:2k_h800_iter_time}
    \end{subfigure}
    \hfill
    \begin{subfigure}[t]{0.32\linewidth}
        \centering
        \includegraphics[width=\linewidth]{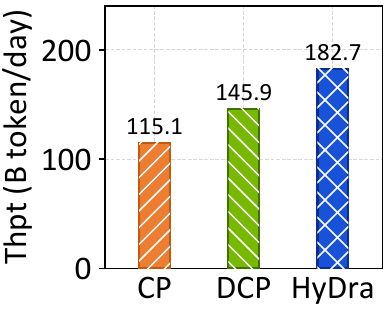}
        \subcaption{\textit{Throughput}}
        \label{fig:2k_h800_throughput}
    \end{subfigure}
    \hfill
    \begin{subfigure}[t]{0.32\linewidth}
        \centering
        \includegraphics[width=\linewidth]{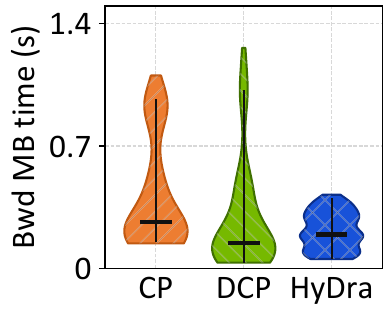}
        \subcaption{\textit{MB-time range}}
        \label{fig:2k_h800_mb_time}
    \end{subfigure}
    \caption{\textit{End-to-end performance on 2,048 GPUs.}}
    \label{fig:2k_h800_performance}
\end{figure}

\textbf{CP communication.}
The degrees the scheduler picks pay off only if the engine can execute them, so we profile one attention layer of an MB under \OURS (Figure~\ref{fig:async_ulysses}).
Both halves of Section~\ref{subsec:comm} behave as intended: the query, key, and value all-to-alls, fully exposed under Mcore's Ulysses (Figure~\ref{fig:ulysses_exposed}), now run behind dependency-free computation, and the outer ring's send/recv pipelines with the attention kernels instead of blocking them.
\OURS thus runs any scheduler-selected degree, including those beyond the head count.

\textbf{Scheduling overhead.}
Because scheduling runs entirely on the CPU, we can measure it at any rank count without the GPUs; Figure~\ref{fig:scheduling_overhead} compares the two schedulers from 2K to 16K GPUs.
Mcore DCP's whole-pool scans reach $47.85$\,s at 16K GPUs, while \OURS stays at $4.12$\,s, an $11.6\times$ speedup.

\textbf{Model convergence.}
\OURS assigns sequences to different ranks and CP-group sizes than static CP, so its gradient and loss reductions accumulate in a different order; because floating-point addition is not associative, a small divergence is expected rather than a defect.
Figure~\ref{fig:loss} bounds it on the aggregate language-modeling loss: over roughly $8{,}000$ iterations, by which point the curve has flattened and the model is close to converged, the two curves stay within $10^{-3}$.
The inset over the final iterations shows them interleaved, and \OURS records the lower loss in around $50\%$ of iterations, so the gap is unbiased noise rather than systematic drift.

\begin{figure}[t]
    \centering
    \captionsetup[subfigure]{skip=-0.5pt}
    \includegraphics[width=0.9\linewidth]{pdfs/eval/pipeline/pp_legend_v2.pdf}\par
    \vspace{2pt}
    \begin{subfigure}[t]{\linewidth}
        \raggedright
        \includegraphics[height=0.185\linewidth]{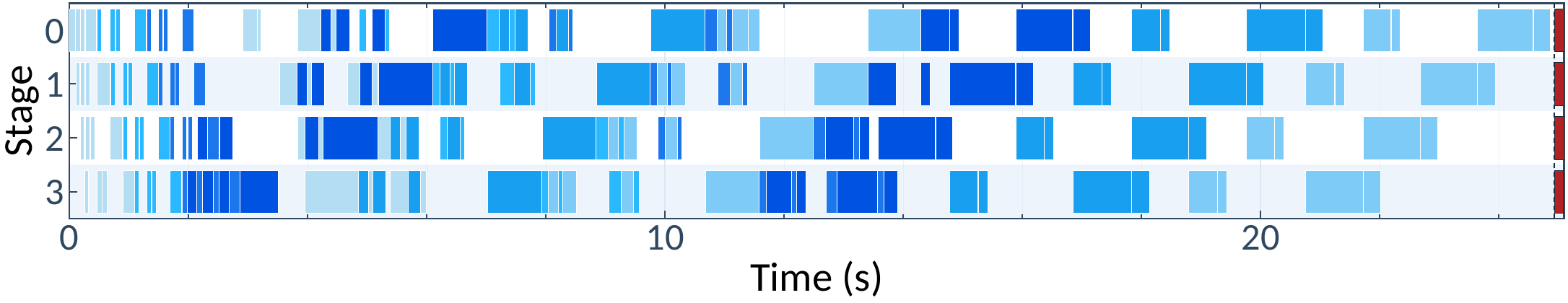}
        \vspace{2pt}
        \subcaption{\textit{Static CP}}
        \label{fig:trace_cp}
    \end{subfigure}
    \begin{subfigure}[t]{\linewidth}
        \raggedright
        \includegraphics[height=0.185\linewidth]{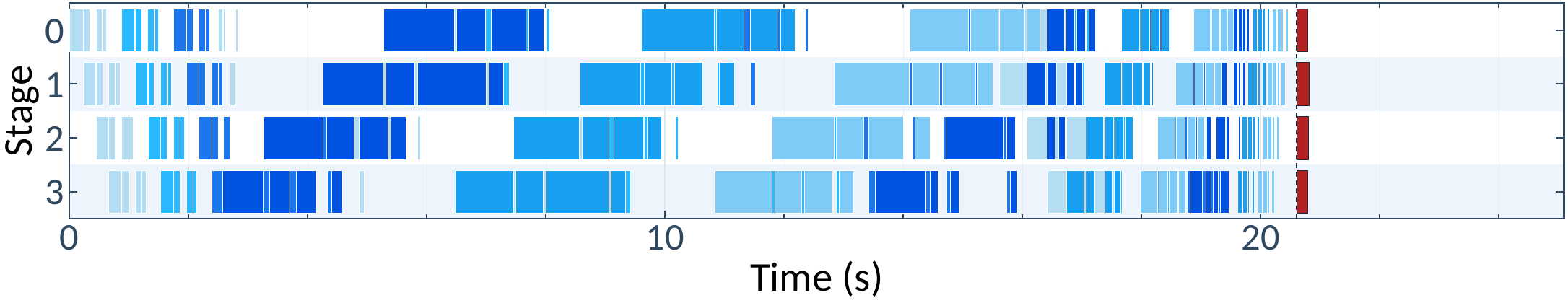}
        \vspace{2pt}
        \subcaption{\textit{Mcore DCP}}
        \label{fig:trace_dcp}
    \end{subfigure}
    \begin{subfigure}[t]{\linewidth}
        \raggedright
        \includegraphics[height=0.185\linewidth]{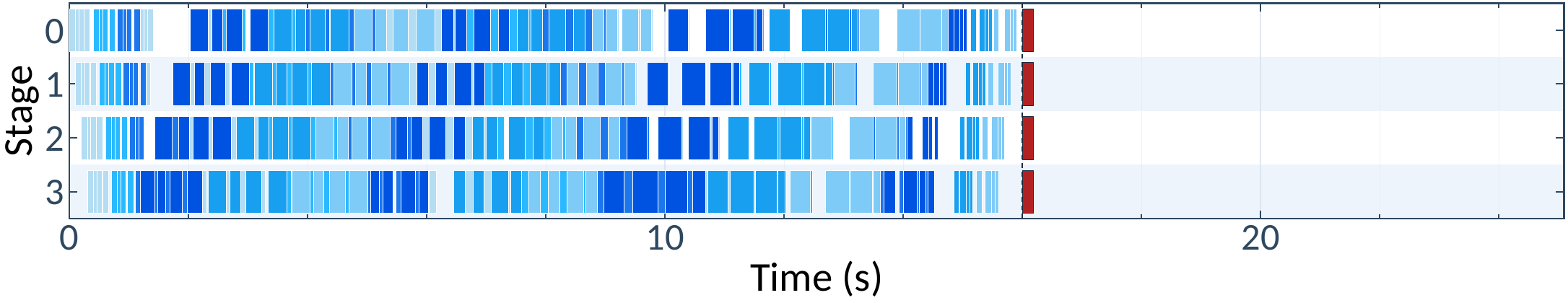}
        \vspace{2pt}
        \subcaption{\textit{\OURS}}
        \label{fig:trace_hydra}
    \end{subfigure}
    \caption{\textit{One iteration timeline across the four PP stages.}}
    \label{fig:2k_h800_traces}
\end{figure}

\begin{figure}[t]
    \centering
    \begin{subfigure}[b]{0.40\linewidth}
        \centering
        \includegraphics[width=\linewidth]{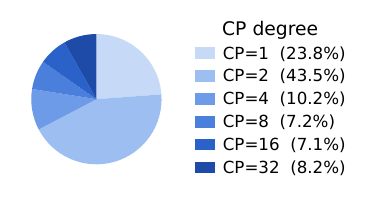}
        \subcaption{\textit{Mcore DCP}}
        \label{fig:cp_gpu_share_dcp}
    \end{subfigure}
    \hfill
    \begin{subfigure}[b]{0.58\linewidth}
        \centering
        \includegraphics[width=\linewidth]{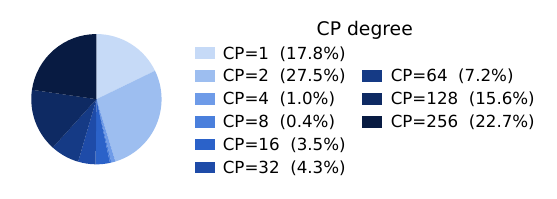}
        \subcaption{\textit{\OURS}}
        \label{fig:cp_gpu_share_hydra}
    \end{subfigure}
    \caption{\textit{Fraction of GPUs at each CP degree.}}
    \label{fig:cp_gpu_share}
\end{figure}

\subsection{Production Speedup}
\label{subsec:production}

\textbf{Throughput.}
On the 2,048-GPU production run, \OURS consistently outperforms both baselines, raising end-to-end training throughput by $1.33$--$1.90\times$ (averaging $1.59\times$) over static CP and by $1.10$--$1.43\times$ (averaging $1.25\times$) over Mcore DCP.
On average, it sustains $182.7$ B tokens per day and cuts the iteration time to $15.9$\,s, from $25.2$\,s under static CP and $19.9$\,s under Mcore DCP (Figures~\ref{fig:2k_h800_iter_time} and~\ref{fig:2k_h800_throughput}).
MB times are also far more even: their coefficient of variation is $49\%$ under \OURS, against $73\%$ under static CP and $108\%$ under Mcore DCP, and the maximum MB time falls as well (Figure~\ref{fig:2k_h800_mb_time}), which directly shrinks the PP bubble examined next.

\looseness=-1 \textbf{PP bubble.}
The pipeline bubble wastes $47.1\%$ of each iteration under static CP and $36.4\%$ under Mcore DCP, but only $13.6\%$ under \OURS (Figure~\ref{fig:2k_h800_traces}).
The mechanism is the one measured on the testbed: \OURS spends context parallelism where it balances MBs, placing a far larger fraction of GPUs at high CP degrees than Mcore DCP (Figure~\ref{fig:cp_gpu_share}).

\looseness=-1 \textbf{DP bubble.}
The DP bubble idles $3.2$--$59.2\%$ of the critical path under static CP and $0.4$--$10.1\%$ under Mcore DCP, while \OURS holds it under $1\%$.
Appendix~\ref{sec:supp-traces} traces both bubbles per stream on rank 0: static CP ends in one long reduce-scatter, Mcore DCP replaces it with gaps recurring inside the step, and \OURS leaves neither.

\textbf{Scheduling overhead.}
On the 2K-GPU production run, \OURS cuts online scheduling time from $0.78$\,s to $0.34$\,s, a $2.3\times$ speedup over Mcore DCP.

\textbf{Training cost.}
The throughput gap also translates into rental cost.
A 2T-token budget\footnote{The 2T-token budget is illustrative and does not reflect any production token count.} at the rates of Figure~\ref{fig:2k_h800_throughput} takes $17.4$ days under static CP, $13.7$ under Mcore DCP, and $11.0$ under \OURS.
At \$2 per GPU-hour~\cite{liu2024deepseekv3}, 2,048 GPUs cost \$1.71M, \$1.35M, and \$1.08M, so \OURS saves \$632K over static CP and \$271K over Mcore DCP.

\begin{figure}[t]
    \centering
    \includegraphics[width=\linewidth]{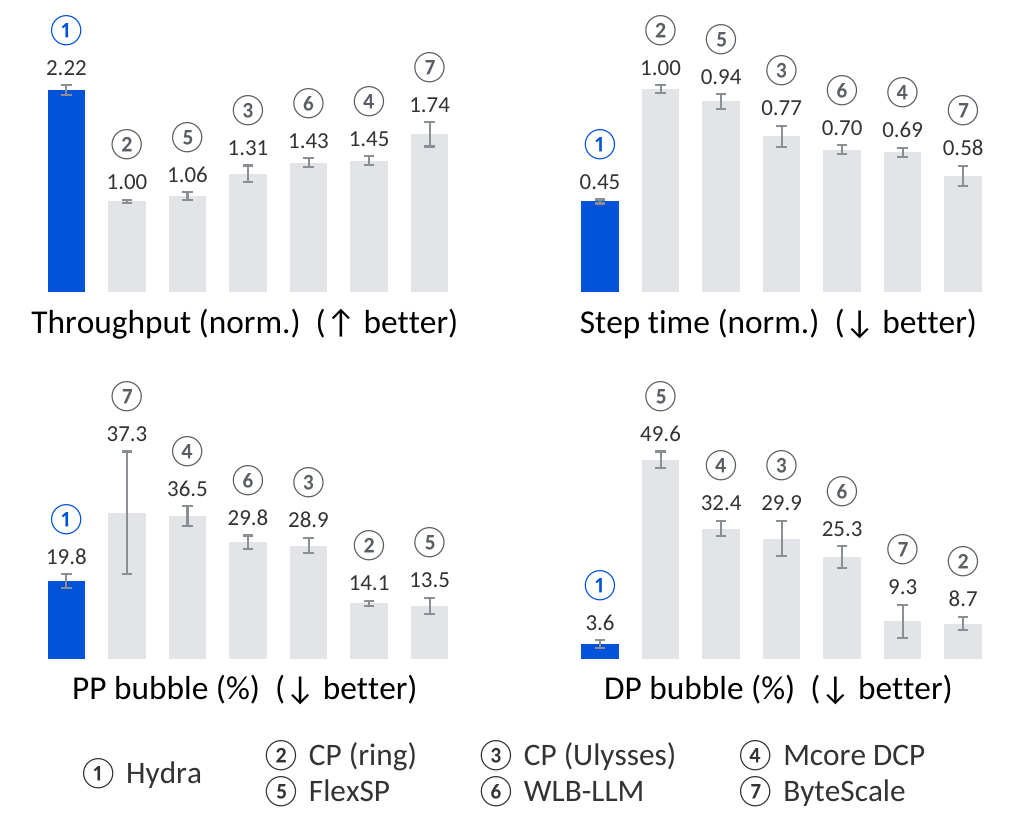}
    \caption{\textit{Simulation results for Hy3 on 8K GPUs; throughput and step time are normalized to static CP (ring).}}
    \label{fig:sim8k}
\end{figure}

\subsection{Scale-Out Simulation}
\label{subsec:simulation}

\textbf{Simulator fidelity.}
Larger scales use an in-house simulator that replays a batch through the full parallel schedule with operator costs measured on the same hardware.
Validated against per-rank production traces at more than 11K GPUs, its steady-state iteration time lands inside the observed range in every configuration and within $10\%$ of the measured mean.

\textbf{Throughput and load balancing.}
Figure~\ref{fig:sim8k} compares the six systems on Hy3~\cite{hy3} at 8K GPUs, where \OURS attains the highest throughput: it improves over the strongest baseline, ByteScale, by $1.28\times$, over Mcore DCP by $1.53\times$, over WLB-LLM by $1.56\times$, and over FlexSP and static CP by $1.7$--$2.2\times$.
The win comes from balance rather than the best value on any single axis: static CP keeps both bubbles low but pays a heavy communication cost, leaving the longest step time; Mcore DCP cuts that cost yet runs high PP and DP bubbles ($36.5\%$ and $32.4\%$); ByteScale trims the DP bubble but leaves the largest PP bubble ($37.3\%$); FlexSP trims the PP bubble but inflates the DP bubble to nearly half the iteration ($49.6\%$).
\OURS avoids static CP's communication cost while holding the PP bubble at $19.8\%$ and the DP bubble at $3.6\%$, so it records the highest throughput.

\begin{figure}[t]
    \centering
    \captionsetup[subfigure]{skip=1pt}
    \includegraphics[width=0.8\linewidth,trim=0 0 220 0,clip]{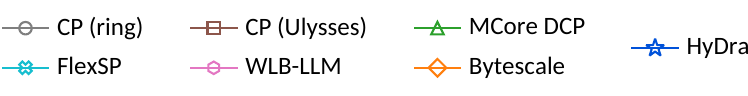}\par
    \begin{subfigure}[b]{0.32\linewidth}
        \centering
        \includegraphics[width=\linewidth]{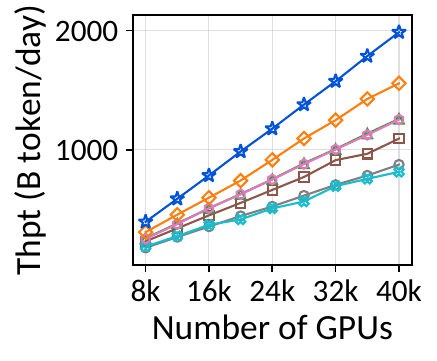}
        \subcaption{\textit{Thpt (Hy3)}}
        \label{fig:scaleout_hy3_throughput}
    \end{subfigure}
    \hfill
    \begin{subfigure}[b]{0.32\linewidth}
        \centering
        \includegraphics[width=\linewidth]{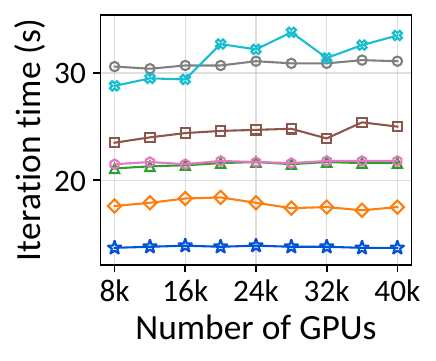}
        \subcaption{\textit{Iter. time (Hy3)}}
        \label{fig:scaleout_hy3_step_time}
    \end{subfigure}
    \hfill
    \begin{subfigure}[b]{0.32\linewidth}
        \centering
        \includegraphics[width=\linewidth]{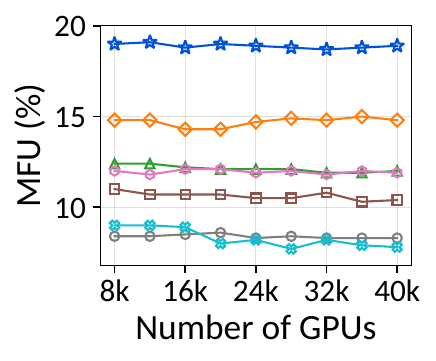}
        \subcaption{\textit{MFU (Hy3)}}
        \label{fig:scaleout_hy3_mfu}
    \end{subfigure}
    \\[4pt]
    \begin{subfigure}[b]{0.32\linewidth}
        \centering
        \includegraphics[width=\linewidth]{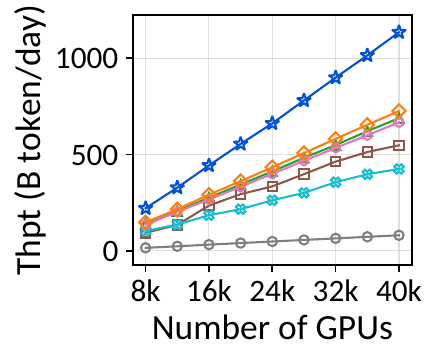}
        \subcaption{\textit{Thpt (DSv3)}}
        \label{fig:scaleout_hy4_throughput}
    \end{subfigure}
    \hfill
    \begin{subfigure}[b]{0.32\linewidth}
        \centering
        \includegraphics[width=\linewidth]{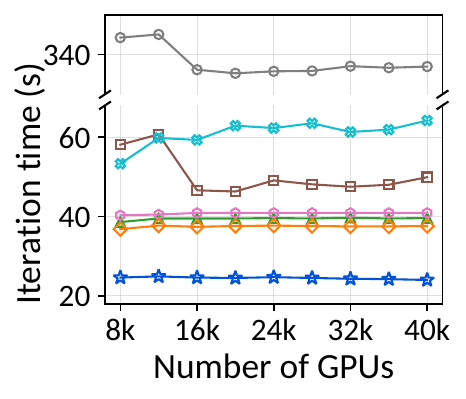}
        \subcaption{\textit{Iter. time (DSv3)}}
        \label{fig:scaleout_hy4_step_time}
    \end{subfigure}
    \hfill
    \begin{subfigure}[b]{0.32\linewidth}
        \centering
        \includegraphics[width=\linewidth]{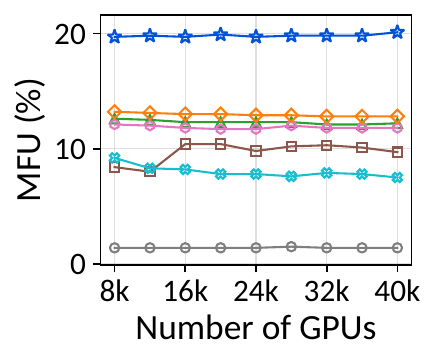}
        \subcaption{\textit{MFU (DSv3)}}
        \label{fig:scaleout_hy4_mfu}
    \end{subfigure}
    \caption{\textit{Weak-scaling performance.}}
    \label{fig:scaleout}
\end{figure}

\textbf{Weak scaling.}
Figure~\ref{fig:scaleout} extends both the GQA model (Hy3) and an MLA model (DeepSeek-V3~\cite{liu2024deepseekv3}) to 40K GPUs, where \OURS stays the fastest at every scale.
On Hy3 the 8K-GPU margins hold across the sweep: averaged over it, \OURS improves throughput by $1.28\times$ over ByteScale, $1.53$--$1.56\times$ over Mcore DCP and WLB-LLM, and $1.7$--$2.2\times$ over FlexSP and static CP.
Its lead is wider on DeepSeek-V3: $1.53\times$ over ByteScale, $1.61\times$ over Mcore DCP, $1.7$--$2.5\times$ over the remaining schedulers, and up to $13.8\times$ over ring static CP, 
whose flat per-rank floor must circulate MLA's full per-head K/V, many times the payload of the eight KV heads Hy3 shares.
\OURS therefore leads on both attention architectures at every scale in the sweep.

\section{Conclusion}
\looseness=-1 We presented \OURS, a scalable load-driven DCP system for long-context training.
Production DCP sizes each CP degree to fit memory, and our study on more than 11K GPUs shows the cost.
Ranks with similar token counts differ by $5\times$ in microbatch time, leaving a $46\%$ pipeline bubble and a $13\%$ data-parallel bubble.
\OURS instead pulls every rank toward one load target, computed in closed form, and places sequences with lazy heaps.
That target asks for larger CP degrees than memory requires, so \OURS nests a Ulysses group in a shallow ring, lowering computation and communication together.
On a \textbf{512}-GPU testbed it raises throughput over Mcore DCP by $1.18\times$ on average at 32K context and $2.48\times$ at 256K, and on a \textbf{2,048}-GPU production job it cuts the pipeline bubble from $36\%$ to $14\%$ and raises throughput by $1.25\times$ on average over Mcore DCP and $1.59\times$ over static CP.
We believe the principles behind \OURS can guide academia and industry in building efficient DCP systems for large-scale long-context training.
\bibliographystyle{ACM-Reference-Format}
\bibliography{sample-base}

\appendix
\section{Rank-0 Execution Traces}
\label{sec:supp-traces}

\begin{figure}[t]
    \centering
    \begin{subfigure}[b]{\linewidth}
        \centering
        \includegraphics[width=\linewidth]{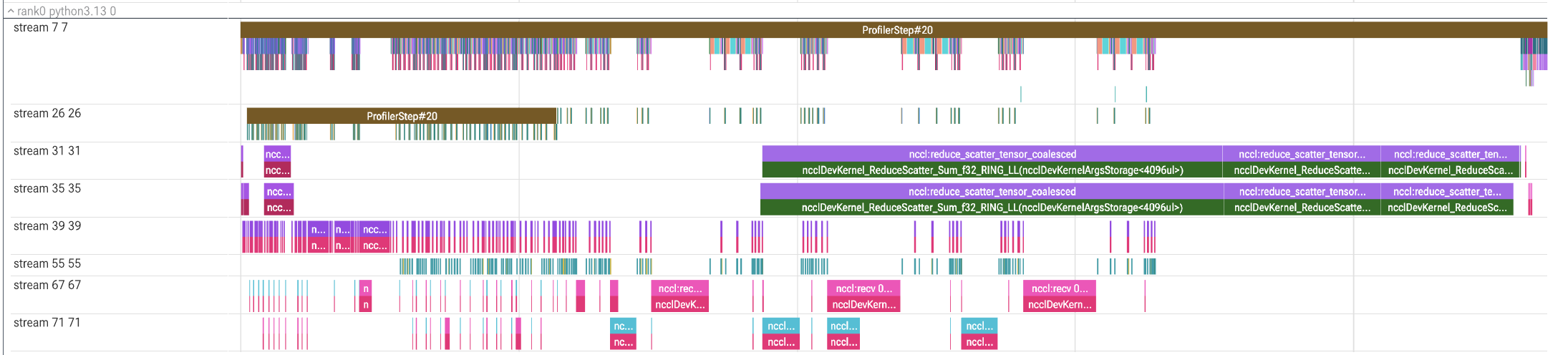}
        \subcaption{\textit{Static CP}}
        \label{fig:rank0_cp}
    \end{subfigure}
    \begin{subfigure}[b]{\linewidth}
        \centering
        \includegraphics[width=\linewidth]{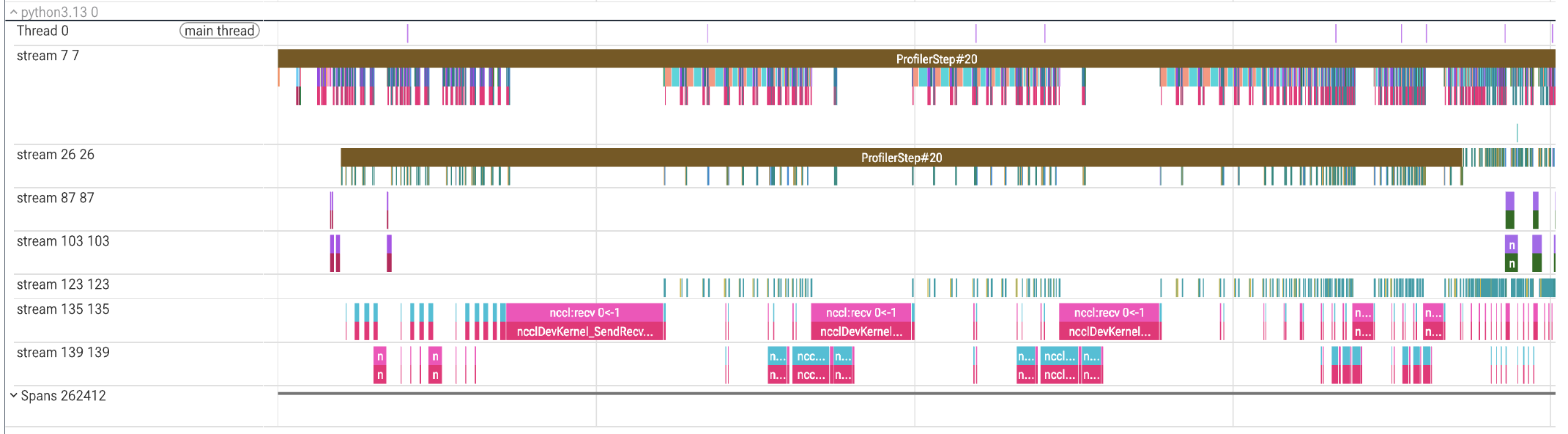}
        \subcaption{\textit{Mcore DCP}}
        \label{fig:rank0_dcp}
    \end{subfigure}
    \begin{subfigure}[b]{\linewidth}
        \centering
        \includegraphics[width=\linewidth]{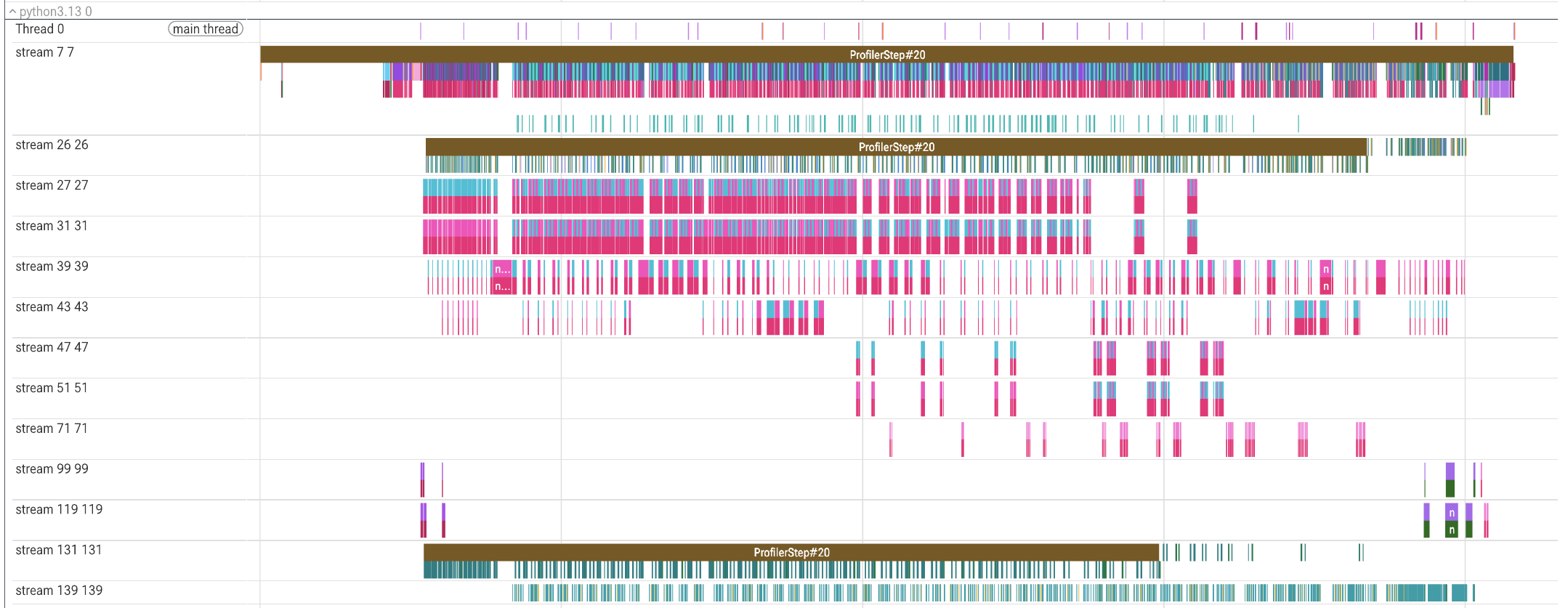}
        \subcaption{\textit{\OURS}}
        \label{fig:rank0_hydra}
    \end{subfigure}
    \caption{\textit{Rank-0 profiler trace of one steady-state iteration under each system.}}
    \label{fig:rank0_traces}
\end{figure}

Figure~\ref{fig:rank0_traces} shows the rank-0 profiler trace of one steady-state iteration of the 2,048-GPU run of Section~\ref{sec:evaluation}, under static CP, Mcore DCP, and \OURS.
Each row is a CUDA stream: the compute stream carries the forward and backward kernels, and the remaining streams carry the NCCL collectives.
Stalls are therefore visible as white space on the compute stream, and the trace localizes them to the barrier that causes them.

\textbf{Static CP stalls at the gradient reduce-scatter.}
The iteration ends in one long collective: the \texttt{ReduceScatter} kernels span roughly the second half of the trace, by which time the compute stream has already gone quiet.
A fixed degree never rebalances computation across sequences, so rank 0 finishes early and idles at the collective until the slowest replica arrives, the DP bubble of Section~\ref{subsec:intramb} at its most severe.

\textbf{Mcore DCP trades it for pipeline idle.}
The closing collective shrinks, but white space appears \emph{inside} the step, where the compute stream alternates between dense bursts of kernels and gaps of comparable width.
A memory-driven degree leaves the heaviest MB heavy and every stage waits on that MB once per microbatch, so these PP bubbles recur through the iteration.

\textbf{\OURS leaves no significant bubble on the compute stream.}
The bursts merge and the wide gaps disappear, leaving only the short intervals between consecutive kernels.
Balancing MBs by attention computation removes the recurring pipeline gaps, and balancing inside each MB keeps the closing reduce-scatter short.

\end{document}